\documentclass[manuscript]{aastex}

\shorttitle{Long Period Giant Extrasolar Planets are Rare}
\shortauthors{Nielsen et al.}

\begin{document}



\title{A Uniform Analysis of 118 Stars with High-Contrast Imaging: Long Period Extrasolar Giant Planets are Rare around Sun-like Stars}


\author{Eric L. Nielsen\altaffilmark{1} and Laird M. Close}
\affil{Steward Observatory, University of Arizona, Tucson, AZ 85721}
\email{enielsen@as.arizona.edu}








\altaffiltext{1}{Michelson Fellow}


\begin{abstract}
We expand on the results of \citet{sdifinalpab}, using the null result for 
giant extrasolar planets around the 118 target stars from the VLT NACO H and 
Ks band planet search \citep{elena}, the VLT and MMT Simultaneous Differential 
Imaging (SDI) survey \citep{sdifinal}, and the Gemini Deep Planet Survey 
\citep{gdps} to set constraints on the population of giant extrasolar 
planets.  Our 
analysis is extended to include the planet luminosity models of 
\citet{fortney}, as well as the correlation between stellar mass and frequency 
of giant planets found by \citet{johnson}.  Doubling the sample size of 
FGKM stars 
strengthens our conclusions: a model for extrasolar giant planets with 
power-laws for mass and semi-major axis as giving by \citet{cumming} cannot, 
with 95\% confidence, have planets beyond 65 AU, compared to the value of 
94 AU reported in \citet{sdifinalpab}, using the models of \citet{cond}.  
When the \citet{johnson} correction for stellar mass (which gives fewer 
Jupiter-mass companions to M stars with respect to solar-type stars) is 
applied, however, this limit moves out to 82 AU.  For the relatively new 
\citet{fortney} models, which predict fainter 
planets across most of parameter space, these upper limits, with and without 
a correction for stellar mass, are 182 and 234 AU, respectively.
\end{abstract}





\section{Introduction}

There are currently close to 300 extrasolar planets known, most detected by 
the radial velocity method \citep{rvref}.  These planets have provided a great 
deal of information on the distribution of giant planets in short period 
orbits.  The likelihood of a star harboring a close-in giant planet 
increases with the metal abundance of the parent star \citep{fv05,santos04}, 
and power laws were found to accurately represent the distributions of mass 
and semi-major axis of exoplanets \citep{cumming}.  While radial velocity 
surveys have moved on to discovering and building up statistics on smaller 
Neptune-mass planets, direct imaging surveys continue to struggle to reach 
even the highest mass planets.

Many observing campaigns have 
been conducted in the last decade to detect and characterize planets through 
direct imaging, especially aimed at young target stars, when the 
self-luminosity of hosted planets is large enough to overcome the glare of the 
parent star.  Improvements in adaptive optics and instrumentation designed 
solely to detect planets, as well as specialized observing techniques, 
have improved the contrasts achievable close to the target star.  This has 
allowed a large increase in sensitivity to planets, and resulted in the 
discovery of several planetary-mass ($<$13 M$_{Jup}$) objects, including 
companions to 
2MASS 1207-3932 \citep{2mass1207}, HIP 30034 (AB Pic) \citep{abpic}, Oph 1622 
\citep{oph1622a,oph1622b,oph1622c}, and DH Tau \citep{dhtau}.  These objects 
were discovered with projected separations of 42, 260, 243, and 330 AU, 
respectively.  Few, if any, of these wide companions ($>$200 AU for objects 
found around stars) are likely to have formed in the primordial 
circumstellar disk of the primary.

Recently there have been exciting 
discoveries of planetary-mass objects around the higher-mass A stars: HR 8799, 
Fomalhaut, and $\beta$ Pic \citep{hr8799b,fomalhautb,betapicb}.  In the case 
of the triple 
planet system HR 8799, we know that the largest separation planet (HR 8799 b, 
with a projected separation of 68 AU) has the same parallax as the primary 
\citep{hr8799laird} and so it (and very likely HR 8799 c and d) formed 
together around the A5 star HR 8799.  While this is an amazing system, in this 
study we concentrate on lower mass stars, more similar to the Sun.

In \citet{sdifinalpab}, we presented null results from the direct imaging 
surveys for extrasolar giant planets of \citet{elena} and \citet{sdifinal}, 
using the contrast curves for each of 60 unique target stars to set 
constraints on the populations of extrasolar planets.  We concluded that 
extrasolar giant planets are rare at large separations ($>$60 AU).  Just 
prior to the publication of our work, \citet{gdps} published the null results 
from the Gemini Deep Planet Survey (GDPS) for 85 stars, reaching conclusions 
very similar to ours.  In this paper, we combine the samples from these 
three surveys, to improve the statistical constraints we can place on 
extrasolar giant planet populations.

Also since these past publications, two important papers have 
been published relevant to direct imaging of extrasolar giant planets.  
\citet{johnson} compared radial velocity target stars of different masses, 
and found that less massive stars have a lower likelihood of hosting a 
giant planet ($>$0.8 M$_{Jup}$).  Since direct imaging surveys lean heavily on 
the M stars in their samples (as these are intrinsically fainter, making the 
detection of close-in planets 
easier), we attempt here to estimate the corresponding decrease in the 
strength of earlier null results.  Also, 
a new set of planet luminosity models have been published by \citet{fortney}, 
which differ from the popular 
``hot start'' models of \citet{burrows} and \citet{cond} 
which had been previously utilized in such work.  These new models are 
based heavily 
on the ``core accretion'' model of planet formation, and tend to predict 
consistently fainter fluxes for giant planets, especially at the youngest 
ages and largest planet masses.  In addition to significantly enlarging the 
sample, this paper takes into account the 
stellar mass dependence of planet frequency, and the new core accretion 
models of \citet{fortney}, to present more realistic constraints on the 
distribution of extrasolar giant planets around Sun-like stars.

\section{Observations}\label{observations}

\subsection{VLT NACO H and Ks imaging}
\citet{elena} carried out a 
survey of 28 young, nearby, late-type stars with the NACO adaptive optics 
system at the 8.2 meter Very Large Telescope (VLT).  These observations have 
exposure times of order 30 minutes, with stars being observed in the H 
or Ks bands.  
For the 22 stars used (see Section~\ref{targetsec}) from the VLT NACO survey of \citet{elena}, the median 
target star is a 12 Myr old K7 star at 30 pc.

\subsection{VLT NACO and MMT SDI}
A survey of 54 young, nearby 
stars of a variety of spectral types (between A and M) was conducted 
between 2003 and 2005, with the results reported in \citet{sdifinal}.  This 
second survey used the Simultaneous Differential Imager (SDI) 
at the 6.5 meter MMT and the 8 meter VLT, an adaptive optics observational 
mode that allows higher contrasts by 
imaging simultaneously in narrow wavelength regions surrounding the 
1.6 $\mu$m methane feature seen in cool brown dwarfs and expected in 
extrasolar planets \citep{lenzen,abdor}  (\citet{swain08} have recently 
detected methane in the atmosphere of the transiting extrasolar planet 
HD 189733b).  This allows the light from a hypothetical 
companion planet to be more easily distinguishable from the speckle noise 
floor (uncorrected starlight), as the two will have very different spectral 
signatures in this region.  This translates to higher sensitivity at smaller 
separations than the observations of \citet{elena}, which were conducted 
before the VLT SDI device was commissioned (see Fig. 14 of 
\citet{sdifinal} for a more detailed comparison of the two surveys).  
For most of these SDI targets, the star was 
observed for a total of 40 minutes of integration time, which includes a 
33 degree roll in the telescope's rotation angle, in order to separate 
super speckles--which are created within the instrument, and so will not 
rotate--from a physical companion, which will rotate on the 
sky \citep{scr1845}.
The 50 stars used from the \citet{sdifinal} SDI survey have a median age, 
distance, and spectral type of 70 Myr, 24 pc, and K1, respectively.

\subsection{Gemini Deep Planet Survey}

At about the same time as the \citet{sdifinal} SDI survey, a direct imaging 
campaign was underway from the Gemini North telescope using the Altair AO 
system and NIRI camera, imaging in a narrow-band H filter with transmission 
between 1.54-1.65 $\mu$m.  The observations were done using the Angular 
Differential Imaging (ADI) technique, which leaves the Cassegrain instrument 
rotator 
off during a sequence of exposures on the star, so that instrumental effects 
like super speckles will stay fixed, 
while physical companions (like planets) will 
rotate throughout the observation \citep{liuadi,adi,gdps}.  This technique is 
most 
effective at producing high contrasts as one moves away from the star, with 
the contrasts achieved exceeding those with SDI \citep{sdifinal} beyond 
$\sim$0.7''.  
For the 71 of the 85 stars from the GDPS survey of \citet{gdps} which we 
consider here, the median target 
star is a K0 star at a distance of 22 pc, with an age of 248 Myr.  Hence, the 
target stars of this survey are somewhat closer and older, whereas the 
southern VLT SDI survey was focused on more distant, though younger, stars.

\subsection{Target Stars}\label{targetsec}

Between the three surveys listed above, our analysis considers 118 distinct 
target stars, with some overlap between surveys.  General properties of the 
target stars, including name, position, distance, spectral type, age, fluxes, 
and observation method are given in 
Table~\ref{table1}.  We attempt to derive 
ages in a uniform manner for all target stars, using the same method as in 
\citet{sdifinalpab}.  If the star is a member of a known moving group, the age 
of that group is adopted as the age of the star.  We limit moving group 
identifications to the well-studied and established groups AB Dor, Her/Lyr, 
Tuc/Hor, $\beta$ Pic, and TW Hya.  Membership in more controversial 
associations, such as the Local Association and IC 2391 (e.g. \citet{laref}), 
are not adopted here.  
If the star is not a member of a group, but has a measured 
value of the calcium emission indicator R'$_{HK}$ and a measurement of the 
equivalent width of the lithium absorption at 6708 $\AA$, the average of the 
ages from the two methods is used.  If only one of these two spectral age 
indicators is available, the age from that measurement is used.  If a star 
from any of the three surveys has none of these three sources for an age 
estimate, it is simply not used in this work.  As a result, 6 stars from the 
\citet{elena} survey, 1 star from the \citet{sdifinal} survey, and 14 stars 
from the \citet{gdps} survey were dropped.

In order to determine ages from the R'$_{HK}$ value, we utilize the 
polynomial fit derived by \citet{Mamajek08}.  The authors derive their 
relation from R'$_{HK}$ values of young clusters, and find a precision of 
0.2 dex for ages derived from this relation.

For lithium values, we compare 
the equivalent width of the 6707 $\AA$ lithium line and effective 
temperature of the star to a set of young stellar clusters.  For each cluster 
(NGC 2264 - 3 Myr \citep{ngc2264_2}, IC 2602 - 50 Myr \citep{ic2602_2}, 
Pleiades - 125 Myr \citep{plelith2}, M34 - 250 Myr \citep{m34}, 
Ursa Majoris - 300 Myr \citep{umalith}, M67 - 5200 Myr \citep{m67}), the mean 
lithium equivalent width is fit as a function of effective temperature.  
Then, for our target stars, we 
interpolate between the fits to each cluster for that star's effective 
temperature, and the lithium value gives us the age (E. Mamajek private 
communication).


For target stars with both a lithium and an R'$_{HK}$ age measurement, the 
median scatter between the two is a factor of 3.  When we consider stars in 
our target list that belong to a single moving group (e.g. AB Dor or $\beta$ 
Pic), and compute their ages using only the lithium or R'$_{HK}$ method (that 
is, we temporarily ignore their membership in a group), we find the scatter 
in the computed age, between members of the same moving group, to also be 
about a factor of 3.  This suggests that the noise in our age measurements is 
primarily astrophysical in nature.
While finding 
a precise age for any single target star is notoriously difficult, our hope 
is that by using a large sample of stars the individual errors will average 
out of our final results.

\begin{figure}
\epsscale{1}
\plotone{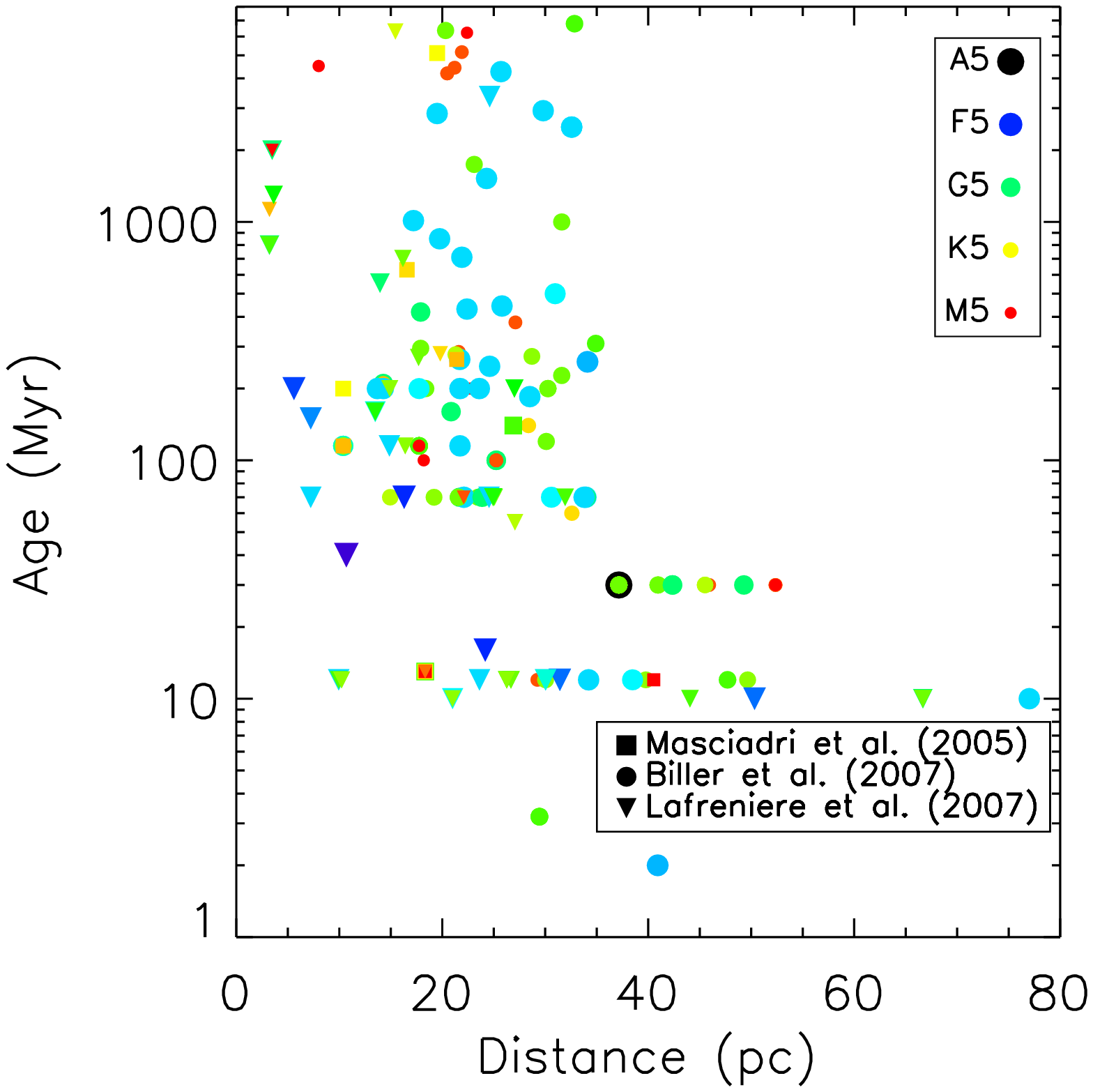}
\caption{The 118 unique stars used in this paper, 
collected from the direct imaging planet 
surveys of \citet{elena} (squares), \citet{sdifinal} (circles), and 
\citet{gdps} (Triangles).  Table~\ref{table1} gives other properties of these 
stars, and Table~\ref{table2} provides details on how the individual ages were 
determined.  The median target star is a 160 Myr K1 star at 24 pc.  The size 
and color of the plotting symbols corresponds to the spectral of 
each target star.  The top legend gives the conversion between size and 
color of the plotting symbol and spectral type: the color scheme follows 
the visible spectrum, with early-type stars represented by large dark purple 
symbols, while late-type stars are small red symbols.
\label{targetsfig}}
\end{figure}

Table~\ref{table2} gives details on measurements (if available) for each of 
the three age determination methods used here, as well as the final adopted 
age, for each target star.  We also plot our targets in Fig.~\ref{targetsfig}, 
giving the age, distance, and spectral type (using absolute H magnitude as a 
proxy) for each star.  Overall, for all 118 of the stars considered in this 
paper, the median target star is a K1 star at a distance of 24 pc with an age 
of 160 Myr.

\section{Monte Carlo Simulations}

As in \citet{sdifinalpab} we use Monte Carlo simulations of ``fake'' planets 
around 
each of the target stars in the three direct imaging surveys considered here.  
A large number (10$^4$ - 10$^5$, depending on the application) of simulated 
planets are given random values of eccentricity, viewing angles, and orbital 
phase based on the appropriate distributions.  Planet mass and semi-major 
axis are assigned either from a grid (see Section~\ref{complete}), or from 
power-law distributions (as in Section~\ref{smastuff}).  For graphical 
representations of the distributions of extrasolar planet orbital parameters, 
see Fig. 2, 5, and 6 of 
\citet{sdifinalpab}, and the discussion therein.  For each observation 
of a given target star, the flux of each simulated planet is computed based 
on the planet's mass and the target star's age, using one of three planet 
models (see Section~\ref{models}).  The angular separation between parent star 
and simulated planet, as well as the flux ratio between planet and star, are 
then computed given the distance to the star.  These are compared to the 
contrast curve for the observation, which give the faintest detectable 
companion (at the 5$\sigma$ level) to the star, in the observation band, as a 
function of angular separation from the star.

In cases where the same target star is observed in multiple epochs, and 
sometimes among different surveys (a common occurrence, the 22, 50, and 71 
stars we use in this work would suggest a sample size of 143 target stars, 
but there are only 
118 unique target stars between these three surveys with reliable age 
estimates), the additional elapsed 
time is taken into account.  Simulated planets are generated at the earliest 
epoch as usual, and compared to that contrast curve.  Their parameters are 
then used again, with orbital phase advanced forward by the time between 
observations (often a small effect for the planets to which these surveys are 
sensitive, a 30 AU orbit around a solar-type star has a 160 year period, 
and the typical time span between observations is at most about 3 years), 
the fluxes of the simulated planets are now computed in the new observation 
band, 
and compared to the new contrast curve.  The process is repeated for as many 
observations as were conducted of the target star, and a simulated planet that 
is detectable in any of the observational epochs is considered detectable.  
Again, \citet{sdifinalpab} provides more details on these simulations, in 
particular their Fig. 3, 4, and 7.

\subsection{Theoretical Models of Giant Planet Fluxes}\label{models}

\begin{figure}
\epsscale{1}
\plotone{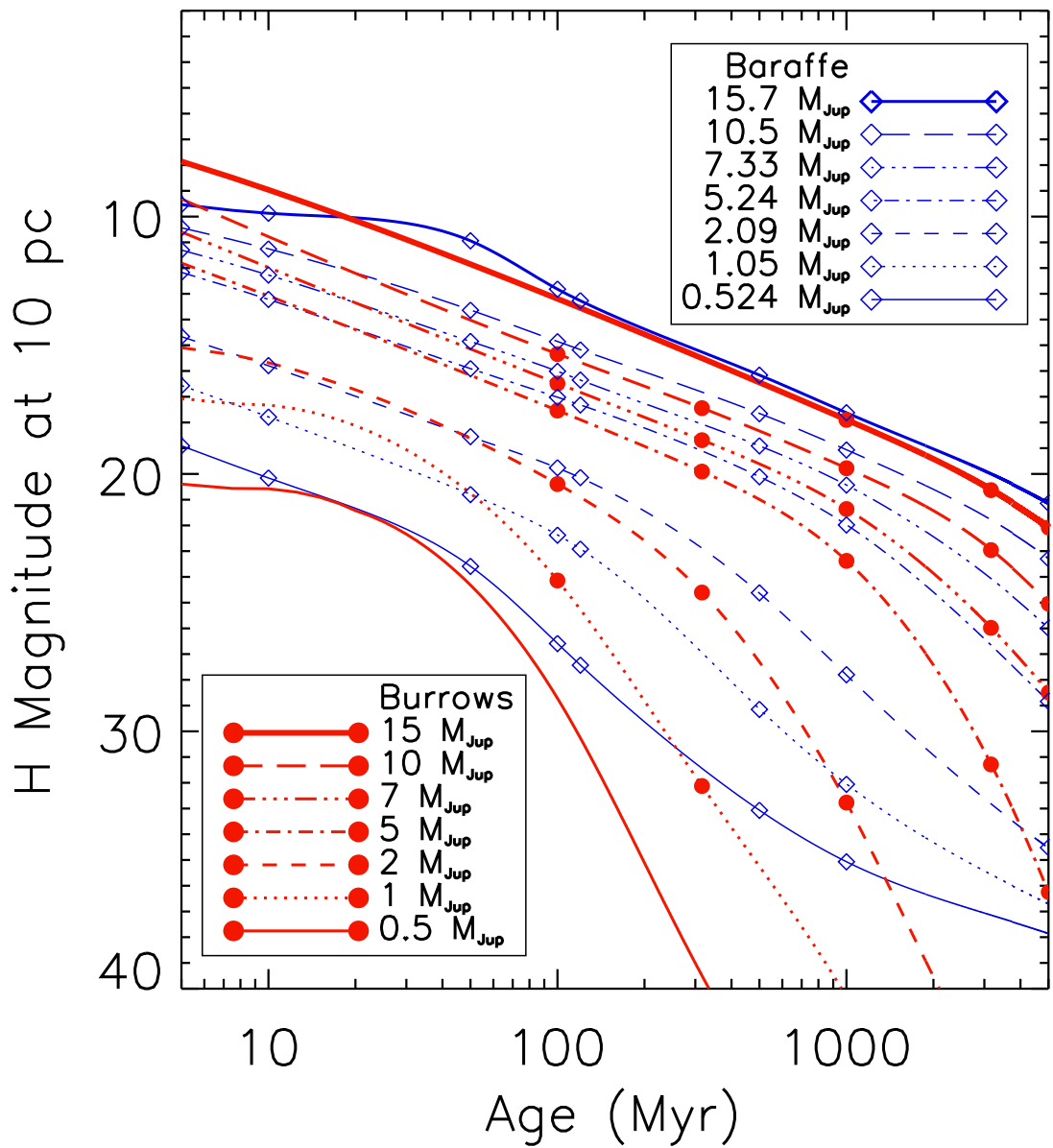}
\caption{A plot of the age and H magnitude of planets, for different masses, 
as predicted by the \citet{cond} and \citet{burrows} models, represented 
by the thin blue lines and the thick red lines, respectively.  The diamonds 
and circles are the H magnitudes given by the models themselves, while the 
lines show the interpolation and extrapolations beyond these points that we 
use when assigning H magnitudes to the simulated planets.  The COND models 
of \citet{cond} required very little extrapolation to fill the range of 
parameter space shown here, while far more extrapolation is required for the 
\citet{burrows} models, especially at young ages and small masses.
\label{modelsfig}}
\end{figure}

In order to use the measured contrast curves for each observed target 
stars to determine which simulated planets could be detected, it is necessary 
to have a conversion from planet mass and age to NIR flux.  As in 
\citet{sdifinalpab}, we use the 
theoretical models of \citet{burrows} and \citet{cond} for the calculation 
of exoplanet flux, using the mass of each simulated planet and the age of 
the host target star, using the filter band (H or Ks) appropriate for the 
particular observation.  In the cases of the GDPS \citep{gdps} and SDI 
\citep{sdifinal} surveys, where the observation band was a specialized filter 
instead of the standard H bandpass, a correction factor is applied 
(see Section~\ref{ntbc_sec} for details).  Though these two ``hot start'' 
models provide basically similar predictions, we perform our calculations with 
both, as the two models can predict significantly different NIR fluxes for 
exoplanets, depending on planet mass and stellar age, as shown in 
Fig.~\ref{modelsfig}.

Since the publication of \citet{sdifinalpab}, an additional set of theoretical 
models have been published by \citet{fortney} for extrasolar planets for a 
range of masses and ages.  The major difference between these new models and 
those from \citet{burrows} and \citet{cond} is that the \citet{fortney} 
models are based heavily on the ``core accretion'' theory of planet formation 
(e.g. \citet{coreacc}), where giant planets are formed from an initial 
$\sim$10 M$_{\earth}$ core accreting gas from the protoplanetary disk.  After 
the brief luminous accretion phase, these 
models predict consistently fainter NIR fluxes than the ``hot start'' models 
(until $\sim$100 Myr to $\sim$1 Gyr, when the models overlap nicely, see 
Fig. 1 of \citet{fortney}), which do not base their initial conditions on 
planetary core accretion models.  For more detail, consult  Figure 8, and 
Tables 1 and 2, of \citet{fortney}.

\begin{figure}
\epsscale{1}
\plotone{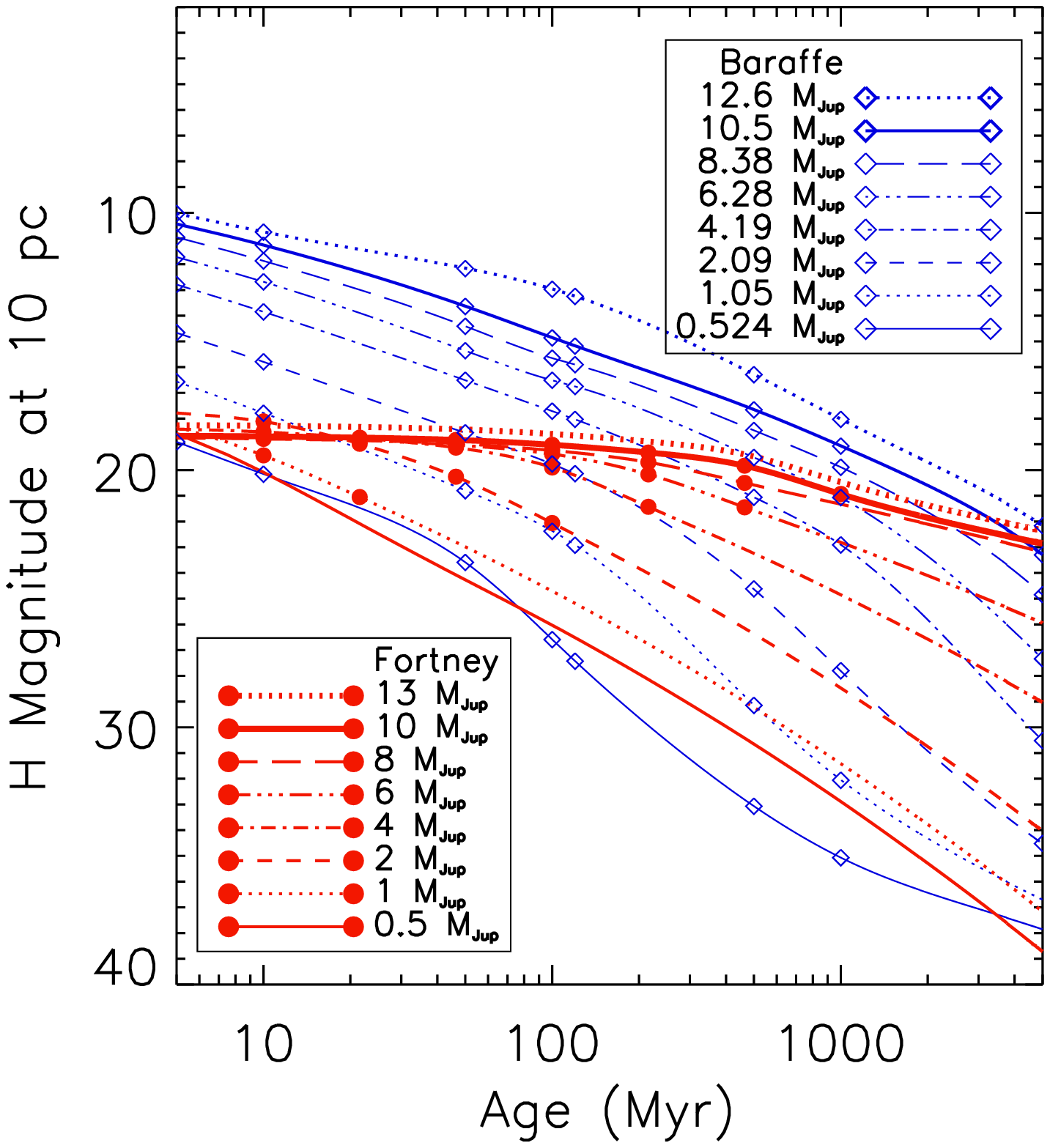}
\caption{As with Fig.~\ref{modelsfig}, a plot of the predicted fluxes of 
extrasolar planets of \citet{cond}, again represented by blue lines and open 
diamonds, this time plotted against the core accretion models of 
\citet{fortney}, the red lines with filled circles.  In order to fill the 
parameter space of planet mass and stellar age we consider, it is necessary 
to extrapolate the \citet{fortney} H magnitudes beyond the grid points of the 
models themselves, especially at larger ages and smaller masses.
\label{modelsfig2}}
\end{figure}

As is the case with the \citet{burrows} models, the \citet{fortney} models 
do not cover the full range of planet parameters we consider here (masses 
between 0.5 and 15 M$_{Jup}$, ages from 1 Myr to 10 Gyr), since 
\citet{fortney} limit their calculations to planets with T$_{eff}>$400K, 
leaving the consideration of cooler planets to future work.  As a result, we 
extrapolate the models to masses below 1 M$_{Jup}$ and above 
10 M$_{Jup}$, and at larger ages (the age a planet cools below 400 K depends 
on the mass of the planet, $\sim$30 Myr for a 1 M$_{Jup}$ planet, and 
$\sim$1 Gyr for a 10 M$_{Jup}$ planet).  While not an ideal solution, as we 
are ignoring the complicated physical processes taking 
places in planets as we cross 
these boundaries in exchange for simple relationships between NIR fluxes and 
age and mass, we believe that overall this method provides a good overall 
picture of the fluxes of extrasolar planets as predicted by the 
\citet{fortney} models.  In Fig.~\ref{modelsfig}, we plot the initial 
gridpoints of both the \citet{cond} and \citet{burrows} models, as well 
as our extrapolations to the full range of parameter space.  A similar plot 
comparing the predicted fluxes for the \citet{cond} and \citet{fortney} 
models is shown in Fig.~\ref{modelsfig2}.  Our effort to 
map additional areas of model parameter space is worthwhile since this work is 
the first to apply these new core accretion models to the field of high 
contrast imaging surveys.

\subsection{Narrowband to Broadband Colors}\label{ntbc_sec}

When we considered stars observed with the SDI method in \citet{sdifinalpab}, 
we used a constant conversion from the broadband H magnitude predicted by the 
models to the measured contrast in the narrowband ``off-methane'' filter 
(SDI F1, 2\% bandpass, centered at 1.575 $\mu$m \citep{abdor}).  While this 
conversion factor was 
consistent with observed T6 objects \citep{sdifinal}, it would be expected to 
vary across a broad range of planet temperatures, corresponding to the large 
differences in ages and masses of the simulated planets.  In this work, we 
used template spectra from the SpeX instrument of 132 low-mass objects, 
spanning spectral types from L0 to T8, to compute the 
difference between broadband H and narrowband 
filters as a function of effective temperature (M. Liu, private 
communication).  Spectral types are converted to effective temperature 
by the polynomial fit of \citet{sptyperef2}, their Table 4.  SpeX spectra 
were obtained from the online SpeX Prism Spectral Libraries (e.g. 
\citet{spex01}, \citet{spex02}, and \citet{spex03}).  Since 
the reliability of the models at reproducing the methane band when modeling 
planet atmospheres is still uncertain, we prefer this method to purely using 
the synthetic spectra from the models to make this color correction.

For GDPS target stars, we use this conversion for the NIRI CH4-short filter, 
to convert the model's prediction of planetary H-band flux to this 6.5\% 
bandpass filter, centered at 1.58 $\mu$m.  For SDI target stars, we follow the 
steps of the data reduction used in computing contrast curves, computing 
planet fluxes for both the bluest ``off-methane filter'' (F1), and the ``on 
methane filter'' (F3) both with a 2\% bandpass, centered at 1.575 and 1.625 
$\mu$m, respectively.  Just as is done for the survey images, the on-methane 
flux is subtracted from the off-methane flux, providing (for each value of 
effective temperature) the expected final flux in the subtracted image, as 
represented by the contrasts curves of \citet{sdifinal}.  For both the SDI and 
GDPS target stars, we use the appropriate effective temperatures predicted by 
the models to match these color corrections to simulated planets of each 
combination of age and mass.

To partially account for this effect in \citet{sdifinalpab}, for SDI target 
stars, we had imposed 
an upper cut-off on planet mass, set by where the models predicted planet 
effective temperatures would rise above 1400 K for a given age.  Above this 
temperature, the methane break would be so weak that subtracting the 
``on-methane'' image from the ``off-methane'' image would simply remove all 
flux from the planet, as it is meant to do for the star.  As a result, 
planets more massive than this limit were simply considered undetectable.  
With our more robust method that appropriately attenuates planet flux as a 
function of temperature, it is no longer necessary for us to impose this 
rather crude binary cut for SDI targets.

In principle, an SDI observation of a non-methanated companion should 
not suffer from self-subtraction of the companion signal, as images in 
the three SDI filters are shifted by wavelength before subtraction.  This 
step aligns the speckles in the images (which scale as $\frac{\lambda}{D}$, 
where $\lambda$ is the observation wavelength and D is the diameter of the 
telescope), but misaligns any physical objects (where separation from the 
primary star on the detector is not a function of wavelength).  As such, 
following subtraction of images from two different filters, a real companion 
should appear as a ``dipole:'' a positive and negative PSF, forming a radial 
line toward the primary star.  The separation between the positive and 
negative parts of the dipole in the subtracted image would be given by 
$\Delta d \sim \frac{\Delta \lambda}{\lambda} d$, 
where $\Delta d$ is the length of the dipole on the detector, 
$\Delta \lambda$ is the difference in wavelength between the two filters, and 
d is the separation on the detector between the primary star and the 
companion.  The most extreme shift in filters for SDI observations is between 
the 1.575 $\mu$m and 1.625 $\mu$m filters, or 3\%.  Since the field of view 
for the NACO VLT SDI observations was only 2.5'', the largest shift between 
positive and negative companions in the subtracted image would be 6.5 
pixels.  As the FWHM for these observations was typically 3.5 pixels, this 
dipole effect would easily be lost against the speckle background at large 
separations, and almost undetectable at small separations, where positive and 
negative companions would more closely overlap \citep{sdifinal}.

\subsection{Completeness Plots}\label{complete}

Using a similar method to \citet{sdifinalpab}, we run Monte Carlo 
simulations of extrasolar planets at a grid of mass and semi-major axis 
points for each target star.  For each star, then, we have what 
fraction of simulated planets could be detected as a function of planet mass 
and semi-major axis.  In order to combine these results over all 118 target 
stars, we again make use of the concept of the ``planet fraction,'' or 
fraction of stars with a particular type of planet, defined such that 

\begin{equation}
N(a,M) = \sum^{N_{obs}=118}_{i=1} f_p(a,M) P_{i}(a,M)
\label{planeteq}
\end{equation}

\noindent where N(a,M) is the number of planets we would expect to detect, as 
a function of semi-major axis and planet mass, N$_{obs}$ is the number of 
stars observed, and P$_{i}(a,M)$ is the fraction of simulated planets, at a 
given combination 
of planet mass and semi-major axis, we could detected around the 
$i$th star in the sample.  $f_p(a,M)$, then, is the fraction of stars that 
have a planet with a mass M and semi-major axis a.  If every star had one 
Jupiter-mass planet at 5 AU, for example, then $f_p(5 AU,1 M_{Jup}) = 1$, and 
the number of these Jupiter analogs we would expect to detect from the three 
surveys would simply be the sum of the detection efficiency for these planets 
around all target stars.  That is, if we had 10 stars in our sample 
($N_{obs}=10$), and we had a 50\% chance of detecting a Jupiter-like planet 
around each star ($P_{i}(5 AU,1 M_{Jup}) = 0.5$), our expected number of 
detections of these planets would be 5.

In the case of not finding planets, as was the case for the three surveys of 
FGKM stars 
considered here, we can use the null result to set an upper limit on the 
planet fraction, $f_p$.  If we assume that planet fraction is constant across 
all stars in our survey (we will reexamine this assumption in 
Section~\ref{johnson}), we can remove $f_p$ from the sum of 
Equation~\ref{planeteq}.  Then, utilizing the Poisson distribution, where 
the probability of 0 detections given an expectation value of 3 
(that is, $N(a,M) = 3$), is 5\%, we 
can set the 95\% confidence level upper limit on planet fraction with the 
equation

\begin{equation}
f_p(a,M) \le \frac{3}{\sum^{N_{obs}}_{i=1} P_{i}(a,M)}
\label{planeteq2}
\end{equation}

\noindent So, with the above example, where the expectation value is 5 for 
Jupiter-analogs over 10 target stars 
($\sum^{N_{obs}=10}_{i=1} P_{i}(5 AU,1 M_{Jup}) = 5$), not detecting any such 
planets would allow us to place a 95\% confidence level upper limit of 60\% 
on the fraction of stars with a Jupiter-twin ($f_p(a,M) < \frac{3}{5}$).  
Doing this over the entire grid of planet mass and semi-major axis allows 
us to plot what constraints can be placed on combinations of these planet 
parameters.

Fig.~\ref{conpaball1} gives the upper limit on planet fraction as a 
function of planet mass and orbital semi-major axis, using the models of 
\citet{cond}, with a similar plot using the theoretical models of 
\citet{burrows} given in Fig.~\ref{conpaball2}.  We can place our strongest 
constraints on planets more massive than $\sim$4 M$_{Jup}$ between 20 and 
300 AU (fewer than 5\% of stars can have such planets at 68\% confidence); 
when stars of all spectral types are considered, the lower limit 
probed by direct imaging and the upper limit of the radial velocity method 
are still a factor of 5 apart.  When we repeat the calculations using the 
models of \citet{fortney}, the decreased NIR flux predicted for giant planets 
reduces constraints that can be placed on extrasolar planets, with the 
``sweet spot'' moving out to $\sim$80 AU, as seen in Fig.~\ref{conpaball3}.

While we continue to run calculations using all three sets of models, and 
report the results here, for the sake of brevity we will henceforth only 
plot figures corresponding to the \citet{cond} COND models.  However, the 
figures 
appropriate to the \citet{burrows} ``hot-start'' and \citet{fortney} core 
accretion models are available 
in our supplement, available at this URL: 
http://exoplanet.as.arizona.edu/$\sim$lclose/exoplanet2.html  The supplement 
also contains individual completeness plots for each of our 118 target stars, 
using each of the three models of planet fluxes.  
Additionally, we summarize basic results for all of our calculations in 
Table~\ref{table3a}.

\subsection{Testing Power Law Distributions for Extrasolar Planet Mass and Semi-Major Axis}\label{smastuff}

These null results for extrasolar planets are also useful in setting 
constraints on the parameters of models for planet populations that assume 
power law distributions for the semi-major axis and mass distributions.  
\citet{cumming} carefully examined the sensitivity of the Keck Planet Search, 
and determined that, over the range to which the radial velocity technique is 
sensitive (0.3 to 10 M$_{Jup}$, 2-2000 day orbital periods), planets follow 
a double power-law distribution with index -1.31 in mass and -0.61 in 
semi-major axis (-0.74 in orbital period).  That is, 
$\frac{dN}{dM} \propto M^{-1.31}$ and $\frac{dN}{da} \propto a^{-0.61}$ (note 
that we define power law indices with respect to linear bins, $\frac{dN}{da}$, 
not the logarithmic bins of \citet{cumming}.  Also, while \citet{cumming} use 
$\alpha$ and $\beta$ to refer to the power law indices for mass and period, 
respectively, we use $\alpha$ to refer to the power law index for semi-major 
axis).

Binarity is likely to disrupt planet formation, or at the very least change 
the underlying distribution of planets between binary planet hosts and single 
stars.  \citet{bonavita07} have shown that the distribution of radial velocity 
planets for binary and single-star hosts are quite similar, and 
\citet{holman99} suggest that planets are stable in binary systems with a 
planet semi-major axis $\gtrsim$20\% of the binary separation.  We take this 
into account for our consideration of power-law distributions of semi-major 
axis by excluding target stars with binaries within a factor of 5 of the 
planetary 
semi-major axis being considered.  In Table~\ref{table3}, we give the results 
of a literature search for binaries among our target stars, including 
binary separation and binary type.

By adopting these power laws, and using the normalization of \citet{fv05} to 
give the total fraction of stars with planets, we can then predict how many 
planets these three surveys should have detected for various power law fits.  
If a large number of planets is predicted, our null result can be used to 
strongly exclude that model.  If we accept the power-law distribution for 
mass of \citet{cumming} and the normalization of \citet{fv05}, the two 
remaining parameters are the semi-major axis power-law index $\alpha$, and the 
semi-major axis upper cut-off (that is, what maximum semi-major axis the 
distribution continues to until planets are no longer present).  We illustrate 
this in Fig.~\ref{assumpfig}, where we depict various models of the semi-major 
axis distribution, and the confidence with which we can reject them, using the 
models of \citet{cond}.  For 12 different combinations of semi-major axis 
power law index and upper cut-off we give the percentage we can reject each of 
these 12 models in this figure.  For the model of 
\citet{cumming}, with $\frac{dN}{da} \propto a^{-0.61}$, and at 95\% 
confidence, the upper cut-off must be less than 65 AU, and less than 30 AU 
with 68\% confidence.

We again use the theoretical models for planet fluxes of \citet{cond}, 
and consider a broader range of power-law index $\alpha$ and upper cut-off 
in Fig.~\ref{smacon1}.  
As before, the results from the two hot start models 
\citep{burrows,cond} are generally similar, as the upper cut-offs must be 
less than 28 and 56 AU (68\% and 95\% confidence) for the \citet{burrows} 
models.  The fainter predicted 
fluxes from the \citet{fortney} models reducing the areas of parameter space 
that our null result can exclude: the 68\% and 95\% 
confidence level upper limits for upper cut-off become 83 and 182 AU, for a 
-0.61 power law index.

\subsection{The Dependence on Stellar Mass of the Frequency of Extrasolar Giant Planets}\label{johnson}

In Sections~\ref{complete} and ~\ref{smastuff}, we assume the distribution 
and frequency of giant planets is constant across all the stars in our 
survey.  \citet{johnson} show this assumption to be incorrect by examining 
the frequency of giant planets around stars in three mass bins from radial 
velocity surveys, and showing that more massive stars are more likely to 
harbor giant planets (see their Fig. 6).  As in \citet{sdifinalpab}, we 
divide the target stars into two samples, one containing only M stars, and 
the other with FGK stars  (Our sample contains a single A star, HD 172555 A, 
with spectral type A5, with all our stars F2 or earlier.  We include this 
A star with the FGK stars; however, observations of this single star are not 
sufficient to make any meaningful statements about the population of planets 
around A stars).  We then imagine a planet fraction ($f_p$) with 
one value for M stars, and another for stars of earlier spectral types.

\begin{figure}
\epsscale{0.85}
\plotone{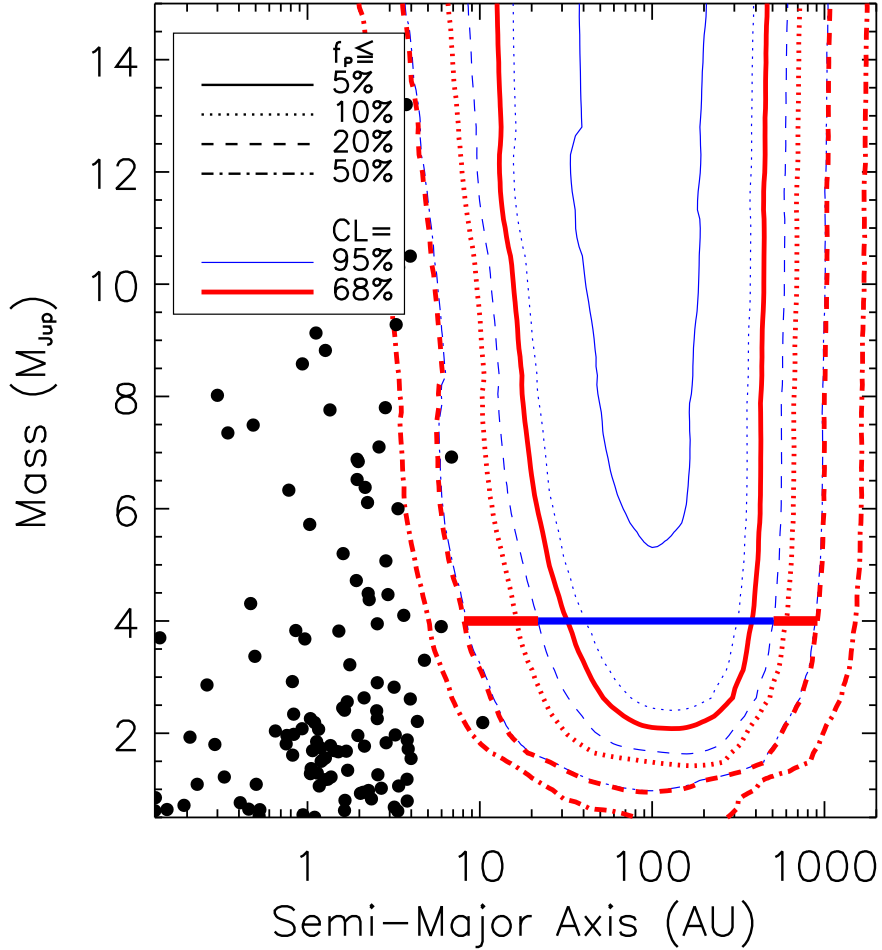}
\caption{The upper limit on planet fraction ($f_p$, the fraction of stars 
with a planet of a given mass and semi-major axis, see 
Equation~\ref{planeteq}), at the 95\% (blue, thin lines) and 68\% (red, 
thick lines)
confidence levels, using the theoretical models of \citet{cond}.  With 
95\% confidence, we can say that less than 1 in 20 stars has a planet 
more massive than 8 M$_{Jup}$ between 50 and 160 AU (constrained by the 
solid blue, thin curve).  We plot a horizontal fiducial bar (again, with a 
thick red line and thin blue line) at 4 M$_{Jup}$, intersecting the 
f$_p \leq$20\% 
contour at both 68\% confidence (outer contour, thick red line) and 95\% 
(inner contour, thin blue line).  Hence, the horizontal line at the bottom 
right of the figure suggests no more than 1 in 5 stars 
would have a planet more massive than 4 M$_{Jup}$ from 8.1 to 911 AU at the 
68\% confidence level, and between 22 and 507 AU at 
the 95\% confidence level.  Known radial velocity 
planets are shown as filled circles for comparison.
\label{conpaball1}}
\end{figure}

\begin{figure}
\epsscale{1}
\plotone{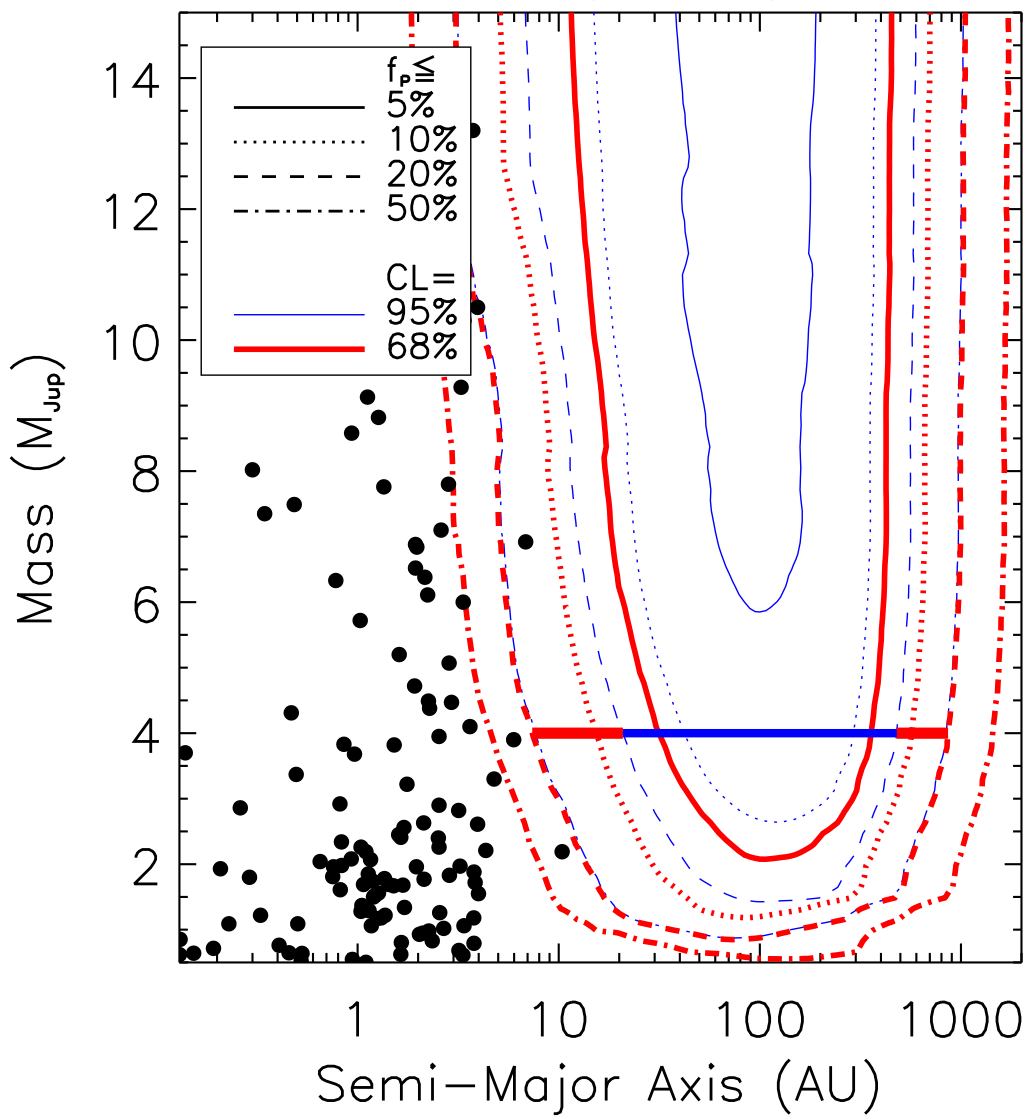}
\caption{As with Fig.~\ref{conpaball1}, the upper limit on planet fraction 
only now using the theoretical models of \citet{burrows}.  The overall shape 
of the graph is quite similar, so with 
95\% confidence, we can place an upper limit on planet fraction of 
5\% for planets larger than 8 M$_{Jup}$ with semi-major axis between 55 and 
130 AU.  As before, radial velocity planets are plotted as solid circles.  
The fiducial 
f$_p \leq$20\% limits for 4 M$_{Jup}$ are between 7.4 and 863 AU at 68\% 
confidence, and 21 to 479 for the 95\% confidence level.
\label{conpaball2}}
\end{figure}

\begin{figure}
\epsscale{1}
\plotone{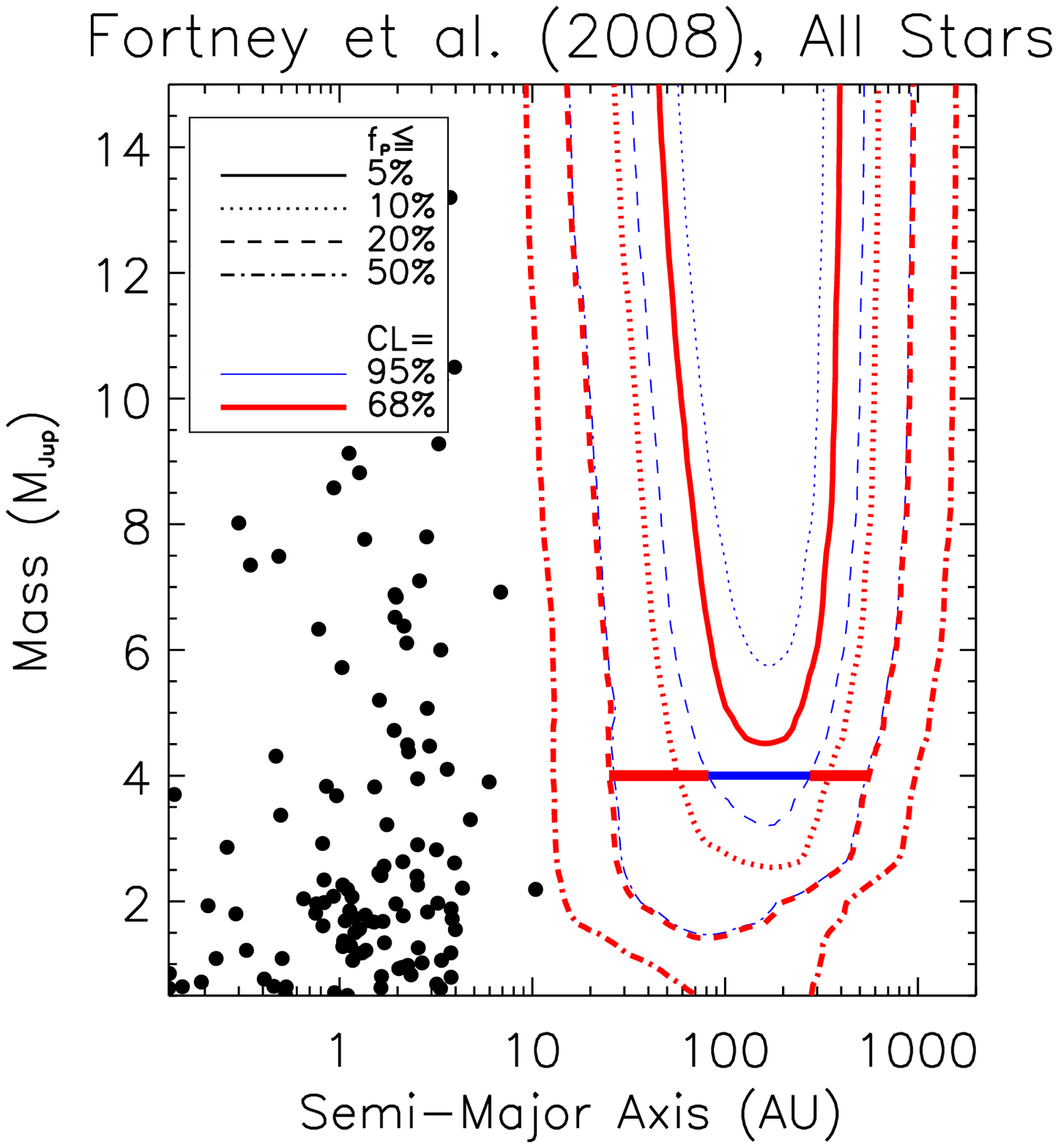}
\caption{The same as Fig.~\ref{conpaball1} and Fig.~\ref{conpaball2}, this 
time using the models of \citet{fortney} to give the upper limit on planet 
fraction.  Overall, the theoretical models of \citet{fortney} are more 
pessimistic as to NIR fluxes of planets when compared to the hot-start 
models \citep{burrows,cond}. With these models, from 82 to 276 AU, less than 
20\% of stars can have a planet above 4 M$_{Jup}$, at the 95\% confidence 
level, and between 25 and 557 AU at 68\% confidence.
\label{conpaball3}}
\end{figure}

\begin{figure}
\epsscale{.8}
\plotone{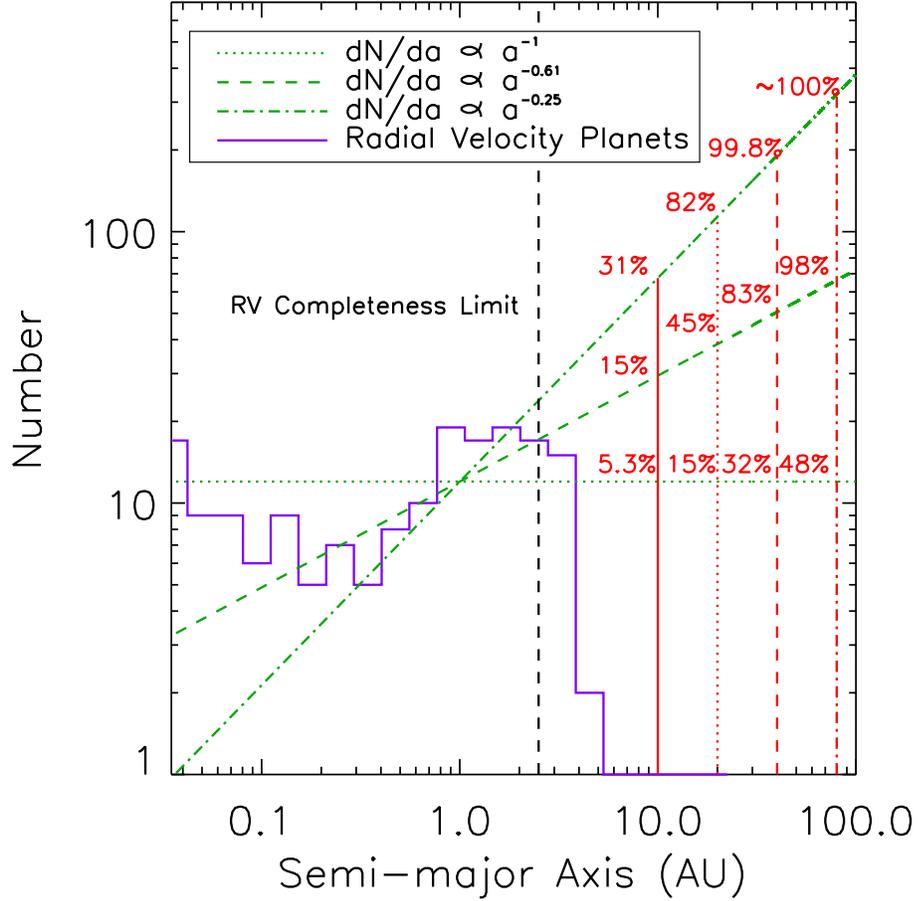}
\caption{Twelve models for the semi-major axis distribution of extrasolar 
planets, using the planet luminosity models of \citet{cond}, with power 
law indices of $\alpha$ = -1, -0.61, and -0.25, and upper cut-offs 
(the limit up to which there are planets, but beyond which planets no longer 
appear) of 10, 20, 40, and 80 AU.  The solid purple line gives the 
histogram of known radial velocity planets, the horizontal and diagonal 
green lines give different values of the power law index, and the red 
vertical lines mark the upper cut-offs.  The vertical black dashed line at 
2.5 AU gives the approximate upper limit to which the radial velocity survey 
is complete to planets.  The percentages at each intersection of power law 
and upper cut-off show the confidence with which that model 
($\frac{dN}{da} \propto a^{\alpha}$ for $a \leq a_{cut-off}$, and 
$\frac{dN}{da} = 0$ for $a > a_{cut-off}$) can be rejected.  For
example, a planet population with dN/da $\sim$ a$^{-1}$ and an outer cutoff
of 10 AU is ruled out at 5.3\% confidence.  For the power 
law of index -0.61 \citep{cumming}, at 95\% confidence the upper cut-off 
must be less than 65 AU, which would fall between the dashed and dot-dashed 
vertical lines of this graph.
\label{assumpfig}}
\end{figure}

\begin{figure}
\epsscale{.65}
\plotone{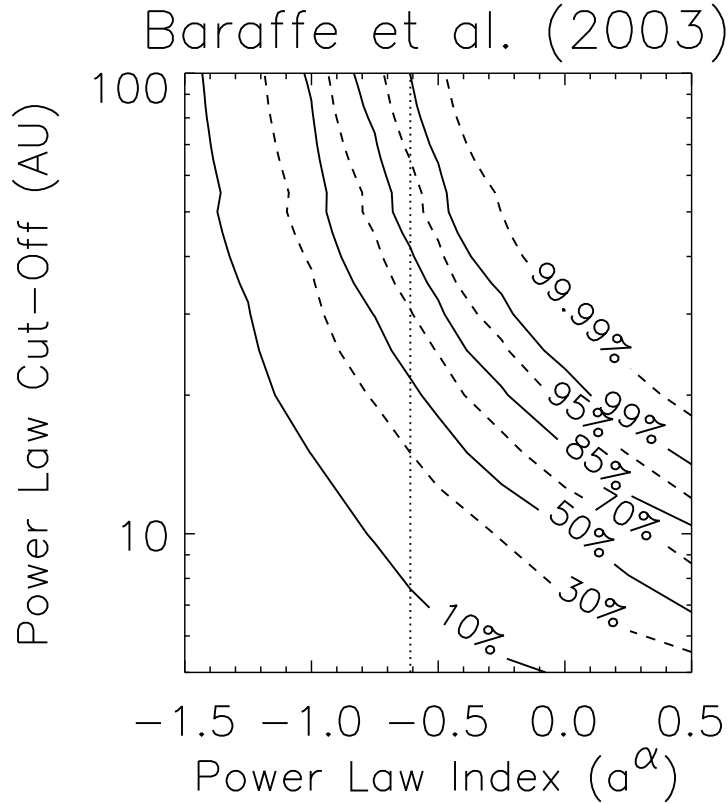}
\caption{Contours showing the confidence with which we can exclude models of 
the semi-major axis distribution of extrasolar giant planets of the form 
$\frac{dN}{da} \propto a^{\alpha}$, with an upper cut-off beyond which there 
are no longer planets, using the models of \citet{cond}.  The power law 
index of -0.61 as given by \citet{cumming} is marked with a dotted line.  The 
jags in the contours are due to binaries being removed as we move up in 
power law cut-off (binary target stars are pulled once the considered 
semi-major axis cut-off reaches one-fifth the binary separation).  The 
pronounced jag between 50 and 55 AU corresponds to the binary M-dwarfs TWA 
8A and TWA 8B (21 pc, 10 Myr) being removed from the sample, indicating the 
strong effect a few M stars have on our results.  For the power law of index 
-0.61, the 68\% and 95\% confidence levels for rejection of this model are 
at 30 and 65 AU.  Similar plots for the \citet{burrows} and \citet{fortney} 
models are available in our supplement, available at 
http://exoplanet.as.arizona.edu/$\sim$lclose/exoplanet2.html ,
 which also contains versions of all future plots 
for these additional two sets of models.  The 68\% and 95\% confidence levels 
are 28 and 56 AU \citep{burrows} and 83 and 182 AU \citep{fortney}.
\label{smacon1}}
\end{figure}

In Fig.~\ref{conpabmstar1} we use the \citet{cond} models to show the 
upper limit that can be placed on the planet fraction for M stars.  Since 
only 18 of the 118 target stars are M stars, the smaller sample size greatly 
reduces the constraints that can be placed on planet fraction near the 
center of the contours, and the outer edge in semi-major axis.  Interestingly, 
the small separation edge of the contours is virtually unchanged between 
Figs.~\ref{conpaball1} and \ref{conpabmstar1}, indicating that the power with 
which these surveys can speak to the populations of short-period giant 
planets is entirely due to the M stars in the surveys.  

Fig.~\ref{conpabfgk1} uses the models of \citet{cond} to give the upper 
limit on planet fraction for the FGK stars in the survey.  The result for 
long-period planets and within the central contours is much the same as for 
stars of all spectral types (Fig.~\ref{conpaball1}), but the contours at the 
smallest values of semi-major axis march outward without the M stars to 
provide high contrasts at small angular separations.  

\begin{figure}
\epsscale{1}
\plotone{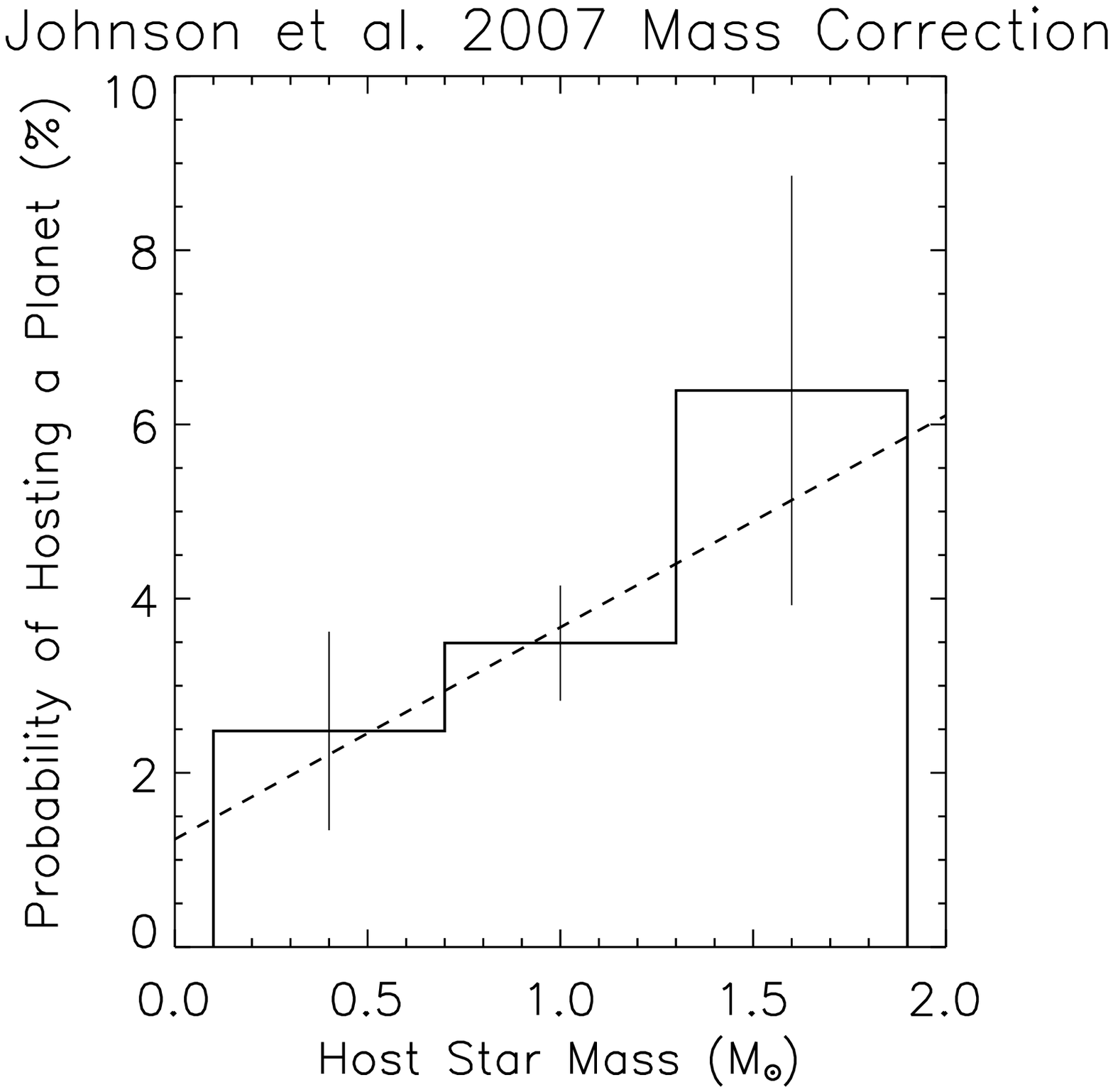}
\caption{Our linear fit to the dependence of the likelihood a target star 
has of hosting a close-in, giant extrasolar planet as a function of stellar 
mass.  The histogram shown is the metallicity-corrected histogram of 
\citet{johnson} (their Fig. 6).  As noted by \citet{johnson}, the probability 
for the high-mass bin is likely underestimated, so future work may show an 
even greater boost for the value of high-mass target stars.  For the target 
stars considered in this work, 35 are in the low-mass bin, 78 are in the 
medium-mass bin, and 5 are in the bin for the highest masses.
\label{mcfig}}
\end{figure}

\begin{figure}
\epsscale{1}
\plotone{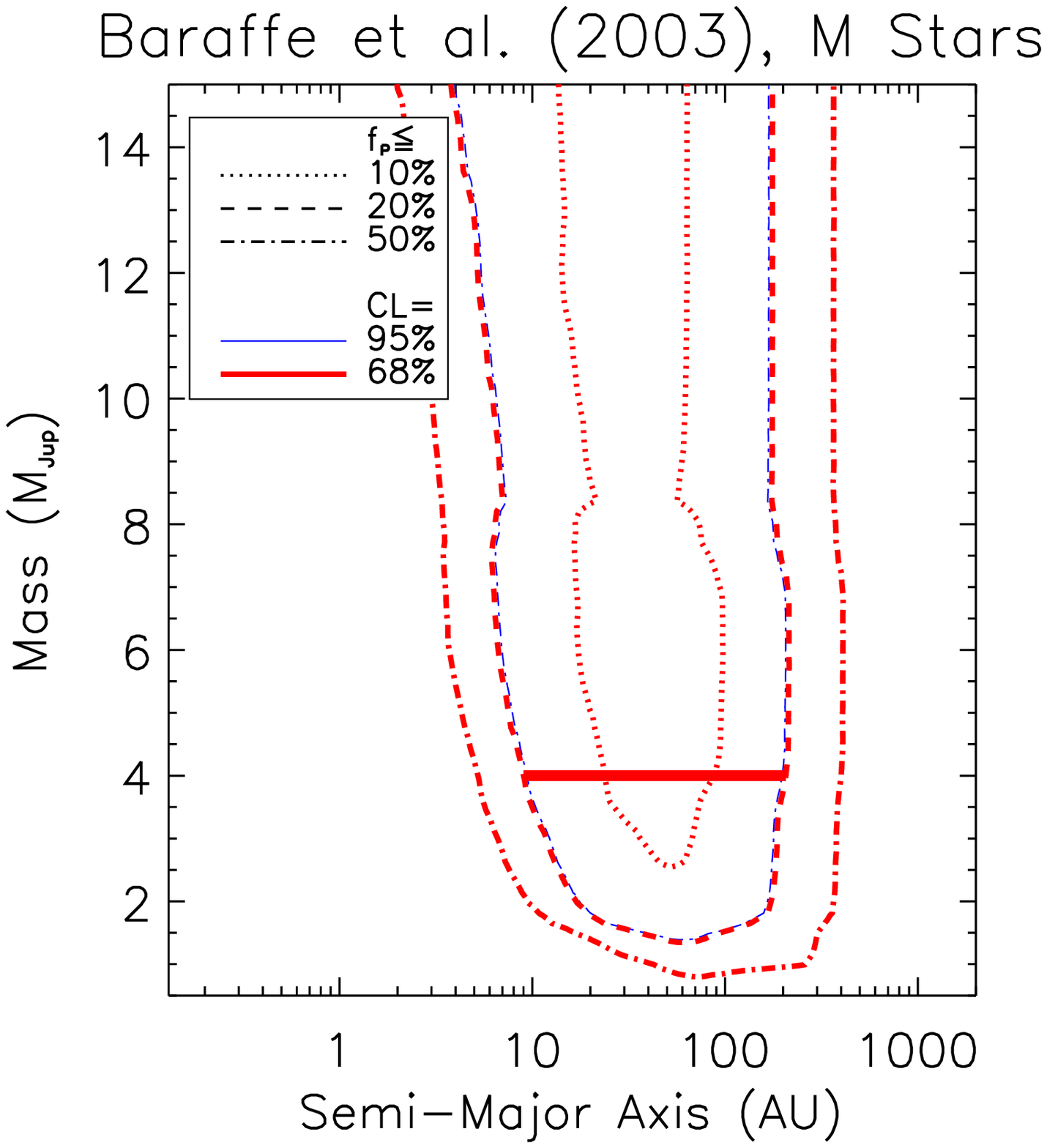}
\caption{Contours giving the upper limit on planet fraction around all the 
stars of M spectral type in the three surveys, using the models of 
\citet{cond}.  Comparing to Fig.~\ref{conpaball1}, which considered stars 
of all spectral types, the behavior of the contours at small semi-major 
axis is roughly the same, while the outer edge and depth of the upper limit 
are limited by the reduced sample size (only 18 of the 118 target stars are 
M stars).  For 68\% confidence, fewer than 1 in 5 stars have a planet more 
massive than 4 M$_{Jup}$ between 9.0 and 207 AU.  For the models of 
\citet{burrows} this range is 8.3 to 213 AU, and is 43 to 88 AU for the 
\citet{fortney} models.
\label{conpabmstar1}}
\end{figure}

\begin{figure}
\epsscale{.9}
\plotone{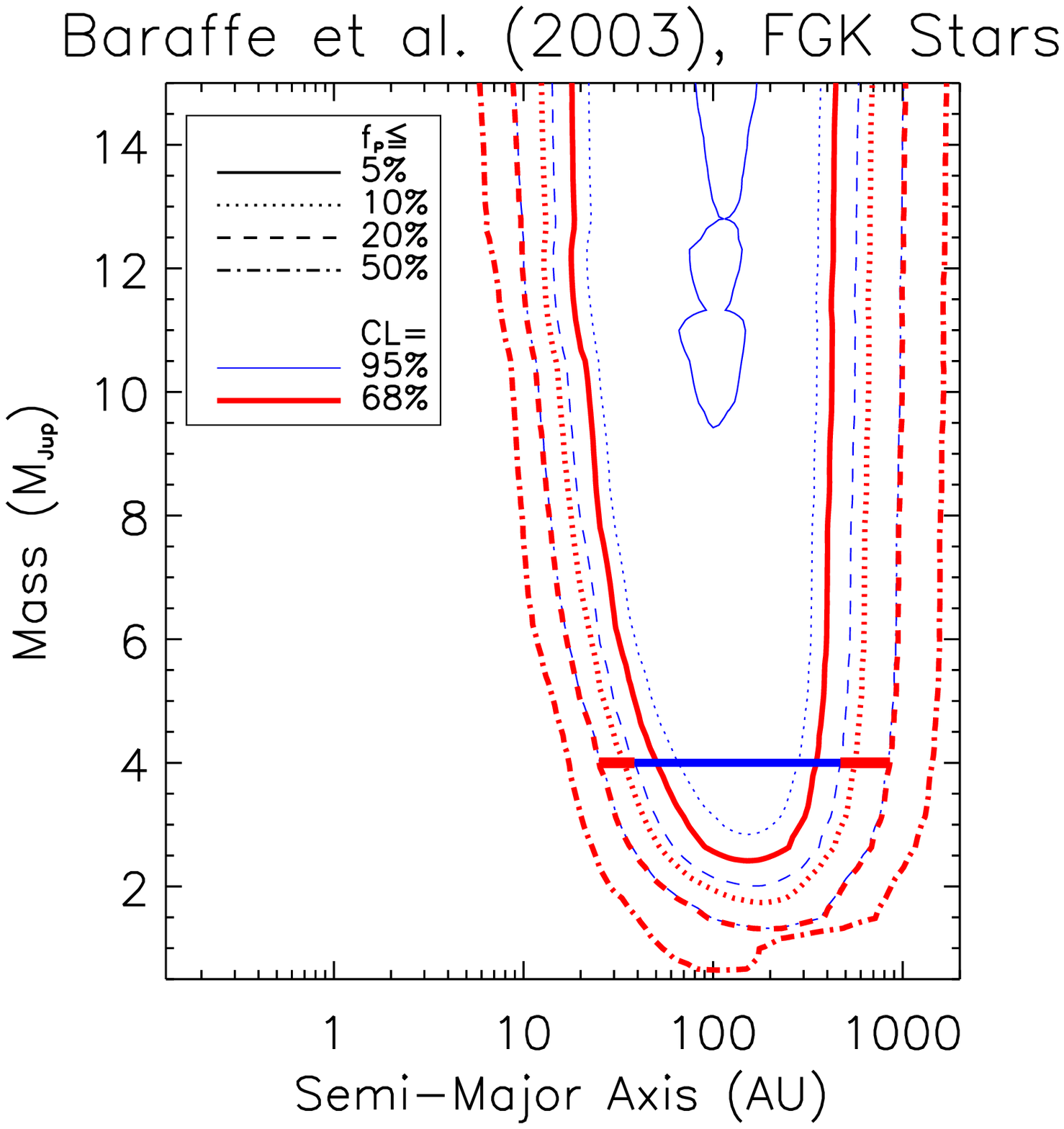}
\caption{The upper limit on planet fraction, using only the FGK stars in 
our survey (as well as the single A star in the survey, HD 172555 A), using 
the models of \citet{cond}.  
The shapes of the contours and the behavior at large semi-major 
axes are roughly the same as in Fig.~\ref{conpaball1}, when all stars 
were considered, but without the M stars and their favorable contrasts at 
small separations, the small-period planets are much less accessible.  The 
20\% contours, at the 68\% and 95\% confidence levels, for planets more 
massive than 4 M$_{Jup}$, are found between 25 and 856 AU and between 38 and 
469 AU, respectively.  For the \citet{burrows} models, the 68\% and 95\% 
limit ranges are between 25 and 807 AU, and 40 and 440 AU; for the 
\citet{fortney} models, the 95\% confidence 20\% contour never reaches 
4 M$_{Jup}$, but the 68\% confidence range is from 59 to 497 AU.
\label{conpabfgk1}}
\end{figure}

A more satisfying way to address the issue of stellar mass dependence is to 
weight the results by target star mass, so that all stars in the survey can be 
applied to the result simultaneously.  To do this, we construct a linear fit 
to the metallicity-corrected histogram from Fig. 6 of \citet{johnson}, to give 
a correction to planet fraction as a function of stellar mass, as we show in 
Fig.~\ref{mcfig}.  (Here we 
assume that the relation found by \citet{johnson} for short-period planets 
(less than six years) applies to the entire range of semi-major axis.  While 
this assumption is obviously untested, in the absence of better data we 
believe it is a good starting point.)  In the 
case of setting upper limits on planet fraction, we now allow planet fraction 
to become a function of stellar mass ($M_*$) in addition to planet mass and 
semi-major axis ($M_p$ and $a$).  In that case we can specify planet fraction 
for the stellar mass of a solar mass ($f_{p,1.0}(a,M_p)$), and find the 
upper limit as with Equation~\ref{planeteq2}, but now including an extra 
term for the mass correction:

\begin{equation}
f_{p,1.0}(a,M_p) \le \frac{3}{\sum^{N_{obs}}_{i=1} P_{i}(a,M_p) mc_{1.0}(M_{*,i})}
\label{planeteq3}
\end{equation}

\noindent where $mc_{1.0}(M_{*,i})$ is the mass correction as a function of 
the stellar mass of the ith star in the sum, normalized to 1.0 M$_{\sun}$, and 
defined by $mc_{1.0}(M_{*}) = \frac{F_{p}(M_*)}{F_p(1.0 M_{\sun})}$, where 
$F_p$ is the fraction of stars with a detected radial velocity planet as a 
function of stellar mass, using the linear fit to the \citet{johnson} 
results.  Again, going back to our earlier example, imagine that we have 10 
stars, each with 50\% completeness to Jupiter-like planets.  If all 10 stars 
are 1 solar mass, then 
$mc_{1.0}(M_{*}) = \frac{F_{p}(1.0M_{\sun})}{F_p(1.0 M_{\sun})} = 1$, and as 
before the upper limit on planet fraction (for the 95\% confidence level, 
as given by the 3 in the numerator) is 60\%.  On the other hand, if only 
four of the ten target stars had masses of 1 M$_{\sun}$, and the remaining 
six had masses of 2.5 M$_{\sun}$, we must weight the results to account for 
the greater likelihood of stars of earlier spectral types to have planets.  
A 2.5 M$_{\sun}$ star is twice as likely to have a planet as a solar mass 
star (see Fig.~\ref{mcfig}), so for the four stars of 
1 M$_{\sun}$, $mc_{1.0}(1.0M_{\sun})$ 
remains 1, as before, while for the stars of 2.5 M$_{\sun}$, this factor 
doubles, $mc_{1.0}(2.5M_{\sun}) = 2$.  In this case, Equation~\ref{planeteq2} 
becomes $f_{p,1.0}(a,M_p) \le \frac{3}{0.5 + 0.5 + 0.5 + 0.5 + 1 + 1 + 1 + 1 
+ 1 + 1} = 3/8 = $37.5\%.  Including A stars in this fictional example almost 
doubles the constraint we can place on the fraction of stars with a giant 
planet.  Similarly, M stars will be weighted against to 
account for their decreased likelihood of having planets.  As an aside, we 
note that while our sample is spread across spectral type (1 A star, 8 F, 33 
G, 58 K, and 18 M stars), only 18 of our target stars are more massive than 
the sun.  Despite the increased probability of finding planets around higher 
mass target stars, these stars are intrinsically brighter, and so moving 
earlier in spectral type very quickly results in any potential planet photons 
being swamped by the glare of its host star (though the recent discoveries of 
planets around A stars, e.g. \citet{hr8799b}, show that this difficulty can 
be overcome and produce exciting results).

\begin{figure}
\epsscale{0.9}
\plotone{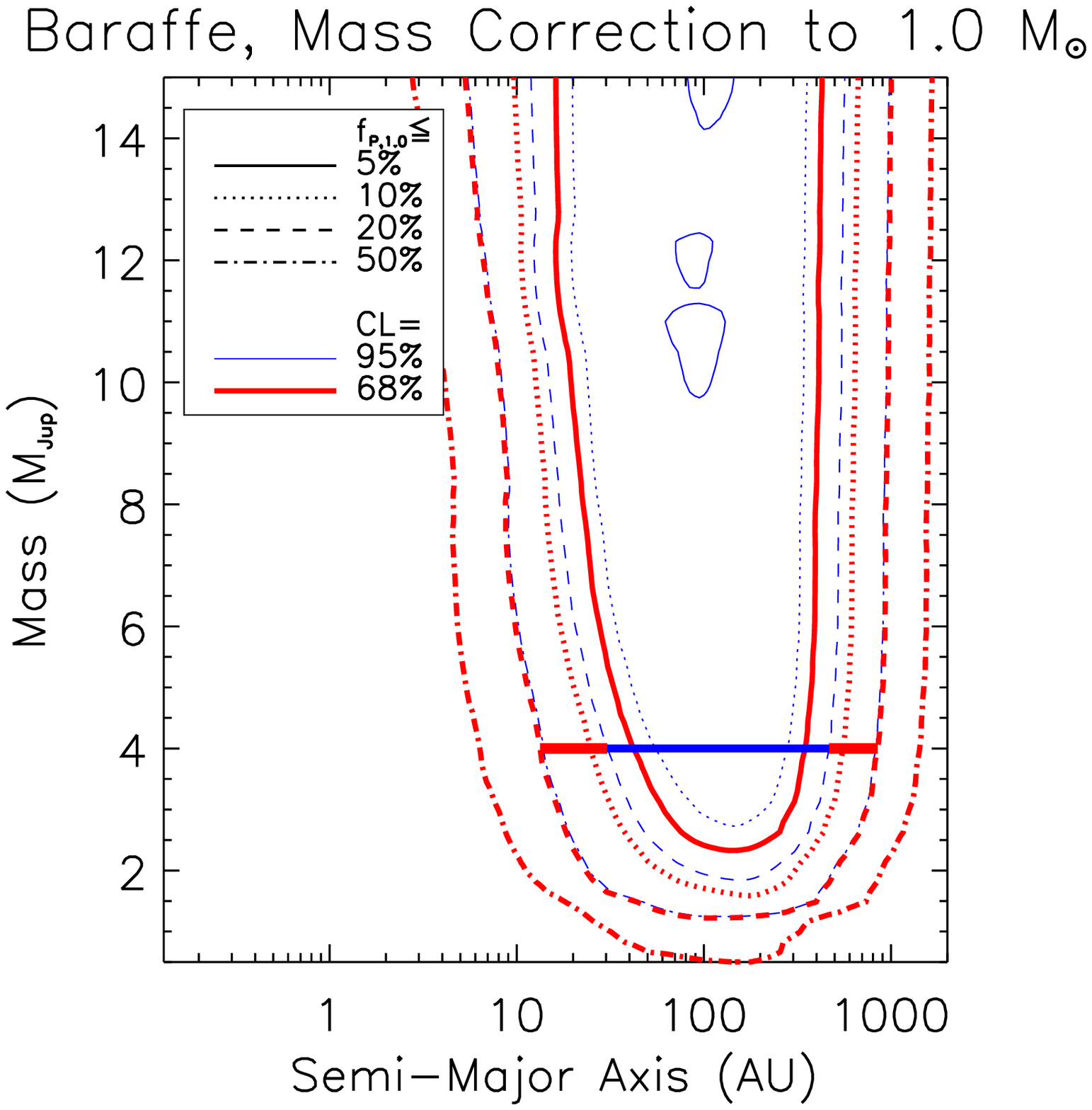}
\caption{The upper limit on planet fraction for stars of 1 M$_{\sun}$ 
($f_{p,1.0}$), with 
constraints from target stars of higher and lower mass stars weighted 
according to a fit to the \citet{johnson} dependence on stellar mass of the 
frequency of radial velocity planets, using the models of \citet{burrows}.  
The plot is similar to that of Fig.~\ref{conpaball1}, which weighted all 
stars equally, but the contours shrink slightly (mainly on the low separation 
side of the plot) as the M stars are now effectively given less 
weight.  The 20\% confidence level for planets more massive than 4 M$_{Jup}$ 
are between 13 and 849 AU at 68\% confidence, and between 30 and 466 AU for 
the 95\% confidence level.  For the \citet{burrows} models, these ranges are 
13 to 805 AU, and 30 to 440 AU, while for the models of \citet{fortney} the 
limits are between 41 and 504 AU, and 123 and 218 AU.
\label{conpabmc101}}
\end{figure}

In Fig.~\ref{conpabmc101}, we plot the upper limit on planet fraction for 
stars of 1 M$_{\sun}$, using Equation~\ref{planeteq3}.  When comparing 
Figs.~\ref{conpabmstar1} and~\ref{conpabfgk1} with Fig.~\ref{conpaball1}, we 
see that the contours at small values of semi-major axis are 
set mainly by the 18 M stars in our sample, while 
the behavior at large separations and the depth of the contours at 
intermediate values of semi-major axis are set by the 100 FGK stars in the 
sample.  So it is then not too surprising that  Fig.~\ref{conpabmc101} is 
quite similar to Fig.~\ref{conpaball1}, with the contours corresponding to the 
smallest upper limits on planet fraction shrinking slightly, and the contour 
at lower semi-major axis moving to the right in the figure, as M stars are 
now given less weight.  

\begin{figure}
\epsscale{0.9}
\plotone{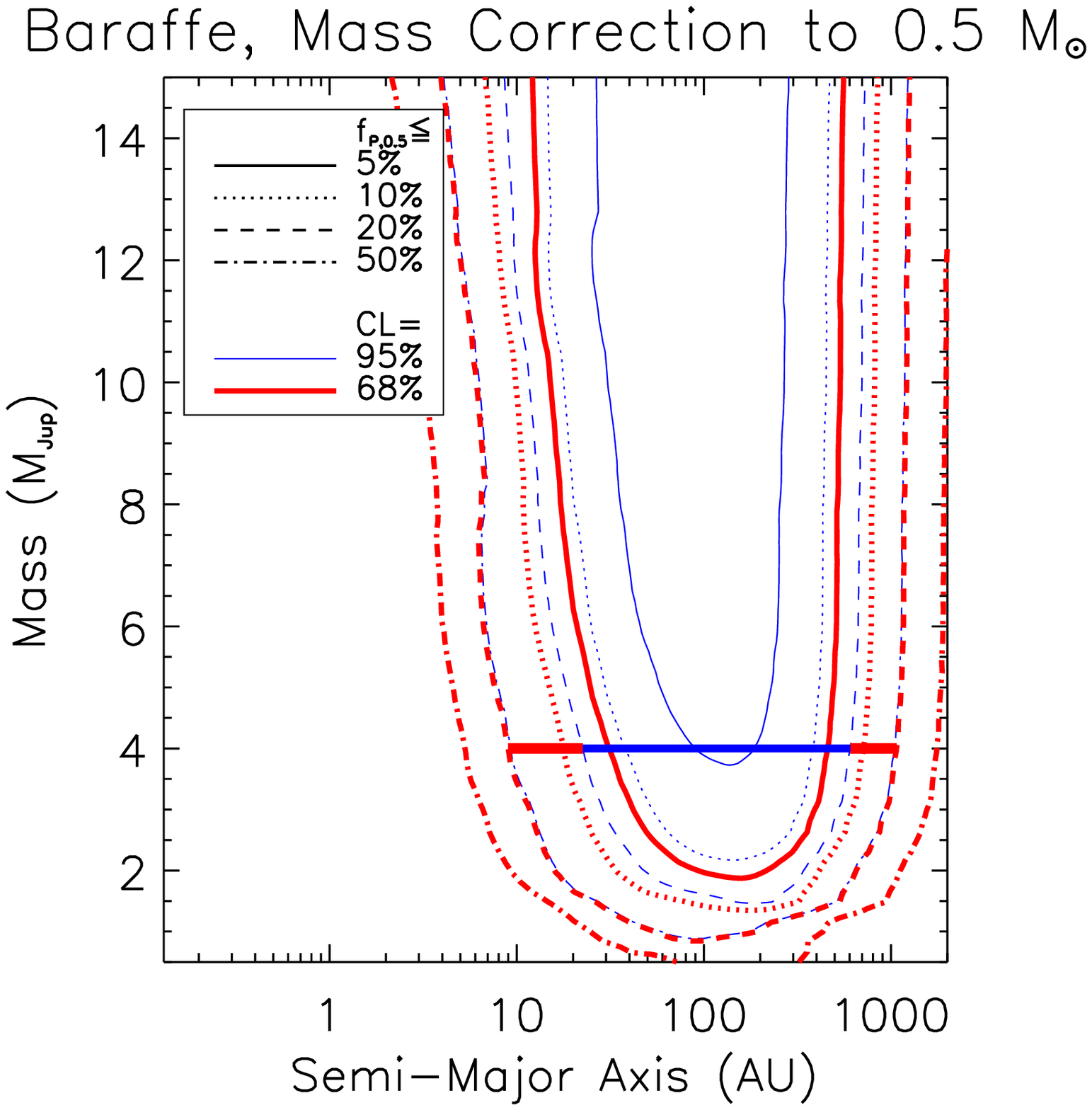}
\caption{The upper limit on planet fraction, this time normalizing to stars 
of a half solar mass (f$_{p,0.5}$), or about M0, using the models of 
\citet{cond}.  
The constraints become stronger, as 
expected, as our assumption going into this calculation is that lower mass 
stars are less likely to have planets overall.  With this set of 
assumptions, the dearth of giant, large-separation planets around M stars is 
made quite clear.  The possibility of lower mass, inner planets around 
M stars (and indeed, planets like those in our own solar system) remains, 
however.  Fewer than 20\% of M stars can have planets more massive than 
4 M$_{Jup}$ between 9.0 and 1070 AU at 68\% confidence, and 23-605 AU at the 
95\% confidence level.  These limits are 8.3 to 1016 AU and from 22 to 573 AU 
for the \citet{burrows} models, as well as 26 to 656 AU and 71 to 341 AU for 
the \citet{fortney}.
\label{conpabmc051}}
\end{figure}

Alternatively, instead of normalizing to solar-type stars, we can instead 
consider what constraints are placed on stars of 0.5 M$_{\sun}$ (about an 
M0 spectral type).  The constraints should become more powerful, as we assume 
a global decrease in the planet fraction for massive planets around lower-mass 
stars.  (Again, this applies strictly to massive planets, $>$0.5$M_{Jup}$.  
The direct imaging surveys considered here are not sensitive to Neptune mass 
planets, which may be more common around M-stars: \citet{endl08} suggest that 
Hot Neptunes may be $\sim$4 times more prevalent orbiting M-stars than Hot 
Jupiters around FGK stars)  In fact, the 
only result of this change is to multiply a constant factor by the 
right-hand-side of Equation~\ref{planeteq3} corresponding to the ratio of the 
likelihood of finding a planet around a solar mass star to that of finding a 
planet 
around a star of 0.5 M$_{\sun}$, or 1.5 in this case.  We plot these limits 
on planet fraction for half solar mass stars in Fig.~\ref{conpabmc051}, with 
the models of \citet{cond}.  As expected, the contours move outward, 
setting strong constraints on the frequency of giant planets in long-period 
orbits around M stars.  

\begin{figure}
\epsscale{1}
\plotone{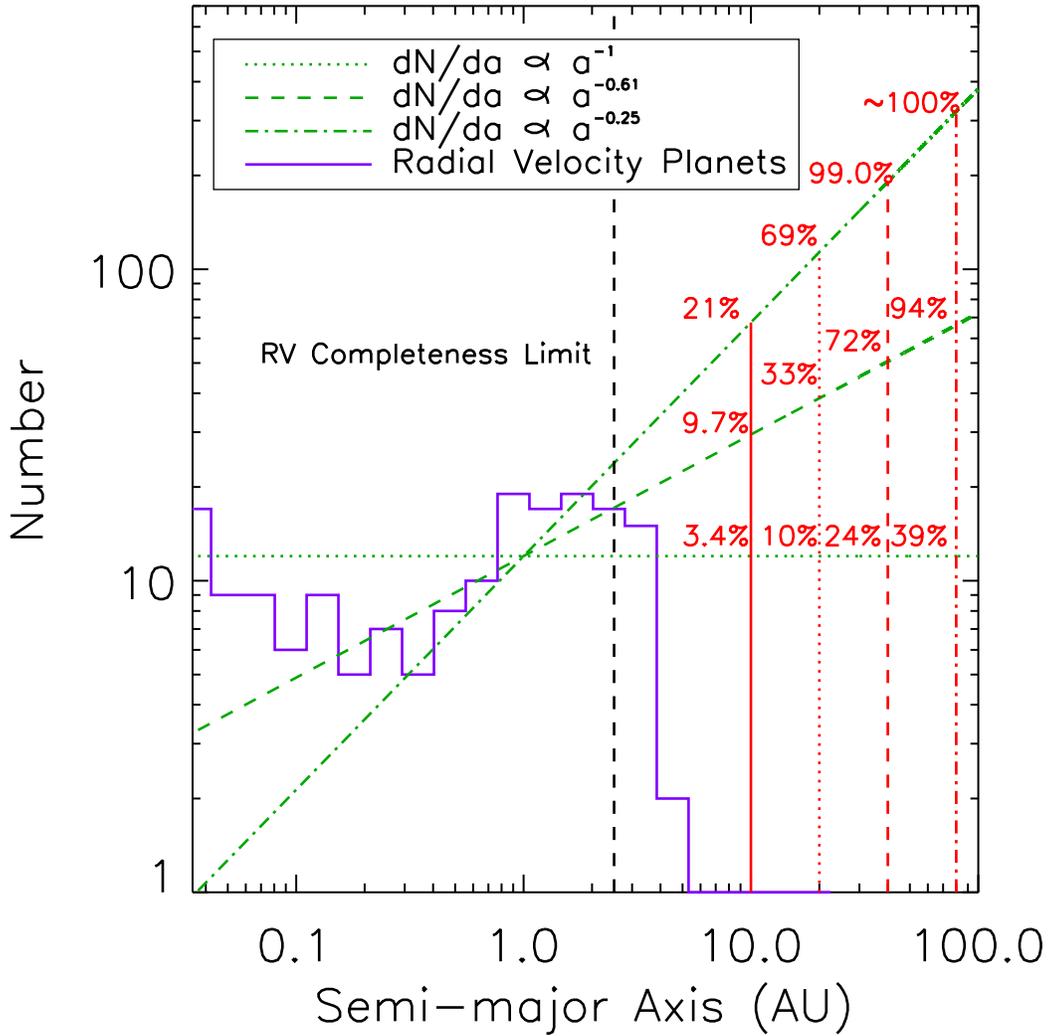}
\caption{As with Fig.~\ref{assumpfig}, power-law models for the semi-major 
axis distribution of extrasolar planets, and the confidence with which we can 
rule out these models, using the results of our survey and the models of 
\citet{cond}.  This time, we utilize the results of \citet{johnson} to 
appropriately give additional weight to high mass stars, which are more 
likely to harbor giant planets.
\label{assumpfigmc}}
\end{figure}

\begin{figure}
\epsscale{1}
\plotone{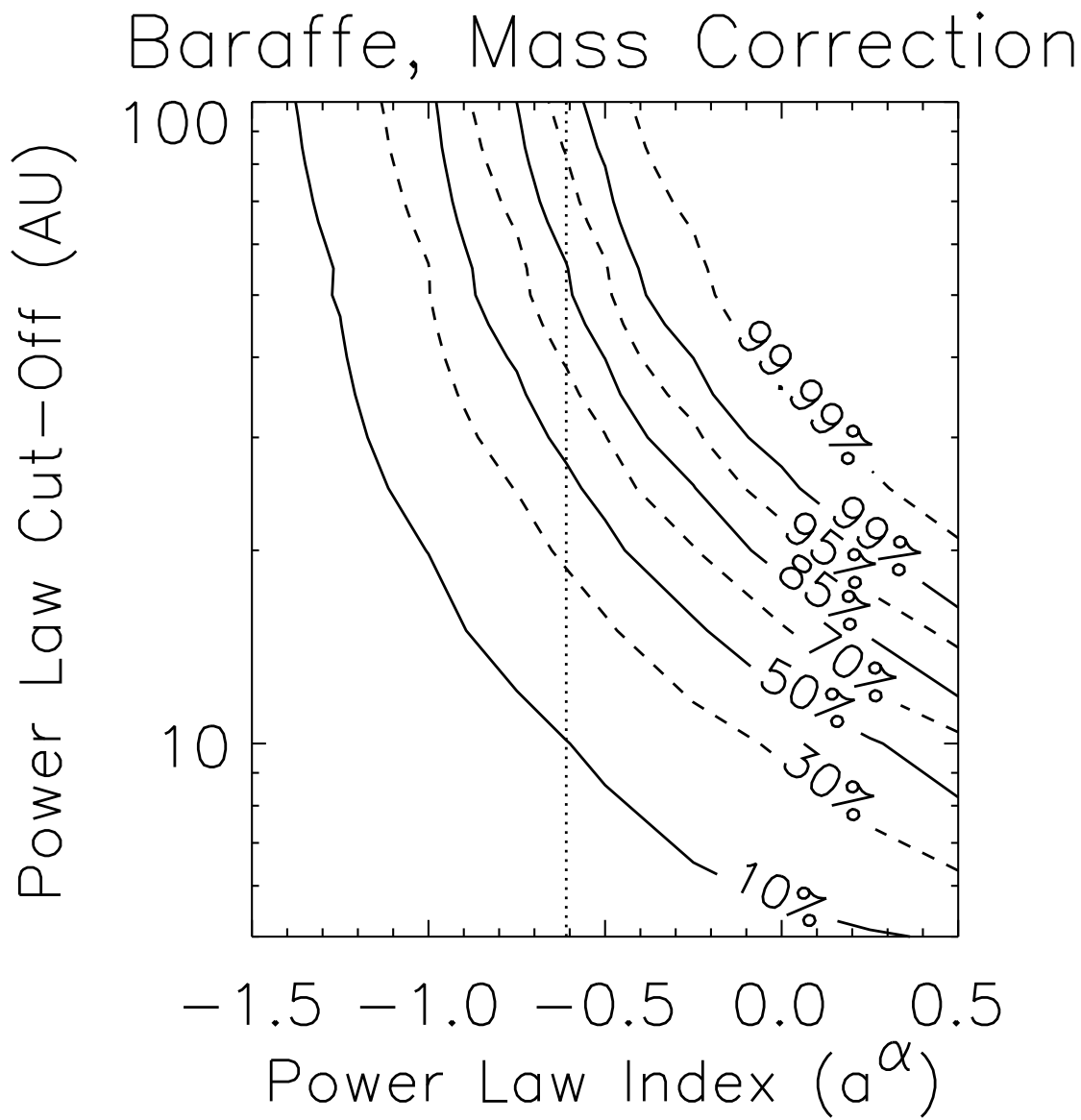}
\caption{Contours showing the confidence with which we can exclude a model 
for the distribution of semi-major axis of extrasolar giant planets given 
by $\frac{dN}{da} \propto a^{\alpha}$ up to some upper cut-off.  This figure 
shows the results for the models of \citet{cond}, with the stellar mass
correction of \citet{johnson} to account for the dependence of likelihood of 
finding giant planets upon the mass of the parent star.  The upper cut-off 
for the \citet{cumming} power law of index -0.61 (as marked by the dotted 
line) is 37 AU at the 68\% confidence level, and 82 at 95\% confidence.  For 
the \citet{burrows} models, the 68\% and 95\% confidence limits are at 
36 and 82 AU, and at 104 and 234 AU for the \citet{fortney} models.
\label{smaconmc1}}
\end{figure}

We also reconsider the implications of stellar mass on the constraints put 
on the power-law model for the semi-major axis distribution of extrasolar 
planets, as discussed in Section~\ref{smastuff}.  Again, by using the linear 
fit to the results of \citet{johnson}, we boost the predicted number of 
planets for higher mass target stars, and suppress that number for lower 
mass stars.  In Fig.~\ref{assumpfigmc} we show the same combination of 
three power law indices and four values of the upper cut-off as before, and 
the models of \citet{cond}, but now with the additional correction for 
the dependence of planet frequency on the stellar mass of each target star.  
The confidence level at which we can exclude each model drops compared to 
Fig.~\ref{assumpfig}, as M stars are effectively given less weight.  While 
the upper limit on planet fraction as a function of planet mass and semi-major 
axis is specific to a given stellar mass, Fig.~\ref{assumpfigmc} (and the 
next one, Fig.~\ref{smaconmc1}) need not be normalized to a specific 
spectral type.  The 
\citet{johnson} mass correction and the \citet{fv05} planet fraction sets the 
absolute likelihood a given target star has a giant planet, which is used to 
calculate the predicted number of planets detected from our entire survey.  
Given our null result, this expectation value is used to set a confidence 
level with which the entire model (giant planet self-luminosity, giant 
planet fraction, dependence of planet fraction on stellar mass, and planet 
mass, semi-major axis, and orbital eccentricity distributions) can be rejected.

For the full range of power law index and upper cut-off, again with the 
\citet{cond} models, we plot contours for the confidence level of 
rejection in Fig.~\ref{smaconmc1}.  This figure is again generally similar 
to Fig.~\ref{smacon1}, but with the constraints slightly looser as M stars 
in the sample receive less weight.  
With the \citet{johnson} mass 
correction, the 68\% and 95\% confidence level upper limits on the semi-major 
axis distribution cut-off are 37 and 82 AU for the \citet{cond} models, 
respectively (without the mass 
correction, these were 30 and 65 AU).  For the \citet{fortney} models these 
move from 83 and 182 AU to 104 and 234 AU.

\subsection{\citet{idalin} Core Accretion Formation Models}

\begin{figure}
\epsscale{0.9}
\plotone{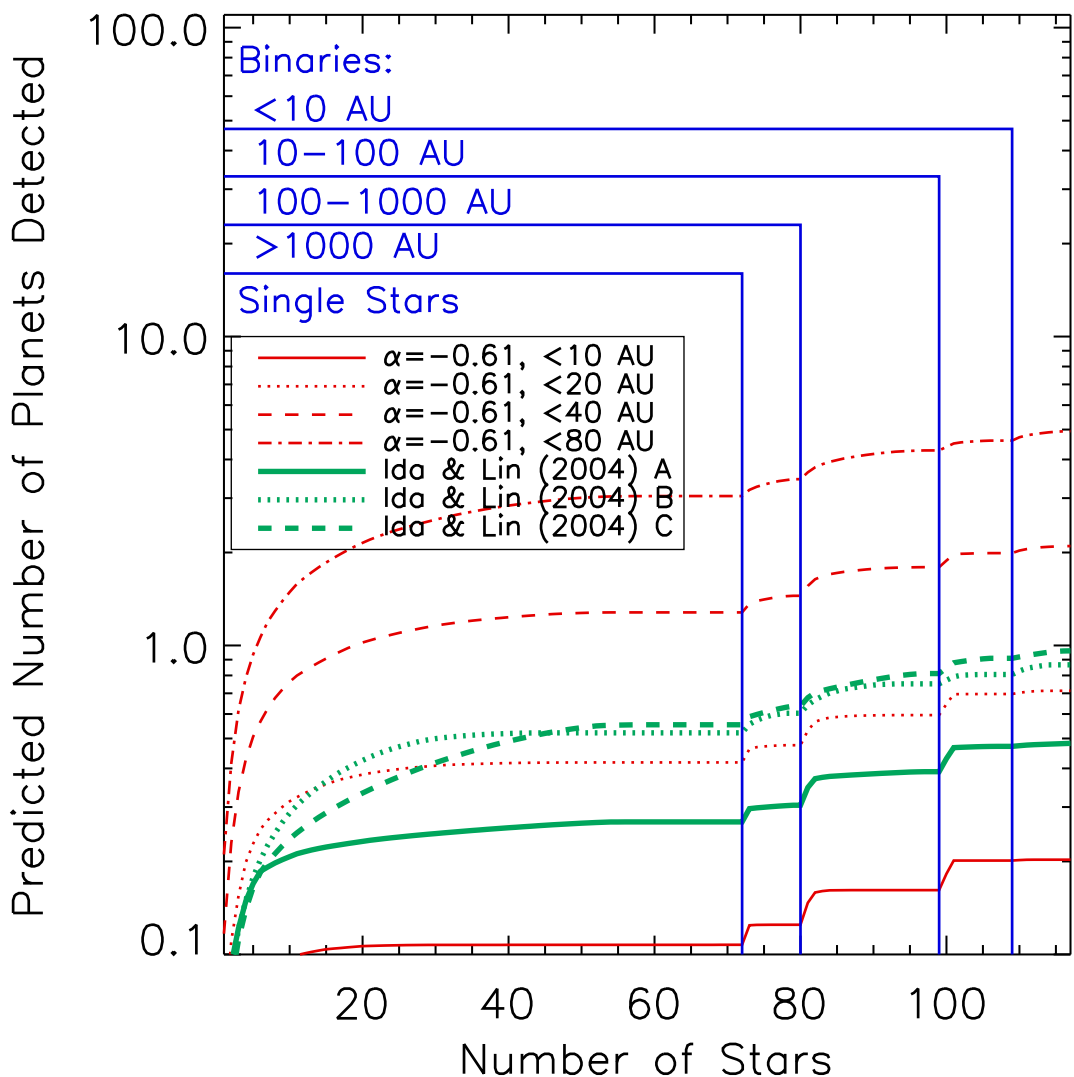}
\caption{The number of planets we'd expect to detect, as a function of the 
number of stars in our survey.  Target stars are divided into bins: one for 
single stars, and binaries divided by separation; within each bin the best 
targets 
are placed to the left of the graph, so they're ``observed first'' in this 
manner.  Using the models of \citet{cond}, and not accounting for stellar 
mass effects or removing binaries, the core accretion models of 
\citet{idalin} predict detecting between 0.5 and 1 planets with the 
combination of the \citet{elena}, \citet{sdifinal}, and \citet{gdps} 
surveys.  The \citet{idalin} models A, B, and C (green curves) can only be 
excluded with 38\%, 58\%, and 62\% confidence, respectively.
\label{surveysize1}}
\end{figure}

As in \citet{sdifinalpab}, we turn to the giant planet formation and dynamical 
evolution models of \citet{idalin}, which predict the final state of giant 
planets, mass and semi-major axis, following the core accretion scenario.  
We extract 200-300 planets from their Fig. 12, and use these masses and 
semi-major axes in our Monte Carlo simulations.  We plot the predicted 
number of planets detected from these models in Fig.~\ref{surveysize1}, 
with target stars divided by binarity.  Even with our 118 target stars, 
without removing close binaries, not accounting for stellar mass (the 
\citet{idalin} models were run with a 1 M$_{\sun}$ primary star), and using 
the planet luminosities of the \citet{cond} models, the Monte Carlo 
simulations show that for each of the three cases of \citet{idalin}, we 
would expect to detect 
about 1 planet for each.  In \citet{sdifinalpab}, we could only exclude the 
cases of \citet{idalin} A, B, and C at 45\%, 49\%, and 50\% confidence, 
respectively, with the expanded target star sample here these rejection 
levels only become 45\%, 59\%, and 63\% (using the \citet{burrows} models to 
be consistent with \citet{sdifinalpab}).  When using the models of 
\citet{cond}, the limits for the sample of this paper are 38\%, 58\%, and 
62\%.  The models of \citet{idalin} 
predict very few giant planets in long-period orbits: fewer than 20\% of the 
giant planets predicted by these models are beyond 10 AU, while our 68\% 
confidence limit on the upper limit of the \citet{cumming} power-law, 23 AU, 
gives 30\% of giant planets in orbits beyond 10 AU.  However, our current 
limits can neither confirm nor rule out the \citet{idalin} populations.

\section{Discussion}

Overall, the conclusions from this work are largely similar to those of 
\citet{sdifinalpab}, that extrasolar giant planets are rare at large 
separations around Sun-like or less massive stars.  Even with the most 
pessimistic models for planet NIR fluxes 
\citep{fortney}, and weighting against the M stars (which provide the most 
favorable contrasts for finding planets), we find that 
at 95\% confidence, fewer than 20\% of solar-mass stars can have a planet 
more massive than 4 M$_{Jup}$ in an orbit between 123 and 218 AU.  Also, a 
power-law model for the semi-major axis distribution of giant planets 
following the results of \citet{cumming} must have a cut-off of 104 AU at 
68\% confidence, and 234 AU at 95\% confidence, again using the 
\citet{fortney} models and the \citet{johnson} mass correction.

It is worth noting that there are additional models of giant 
planet distributions beyond what we consider here.  \citet{cumming} found a 
good fit to distributions of close-in giant planets from radial velocity 
work using a single power-law for the distribution of semi-major axis.  It is 
also possible, however, that while this is a good fit for giant planets within 
$\sim$5 AU, it may not hold for planets at larger separations; perhaps a 
broken power law or some other distribution governs the population of giant 
planets in long period orbits.  If all giant planets are formed beyond the 
snow line, perhaps the final distribution differs for planets that migrate 
inwards and those that remain beyond the snow line.  Alternatively, there may 
be multiple methods of planet formation, such as the core accretion scenario 
and the disk instability model (e.g. \citet{idalin} vs. \citet{boss07}), 
and the frequency with which each occurs is a function of distance from the 
star.  Another possibility is that mass and 
semi-major axis are not independent distributions, which should become 
testable as the number of known exoplanets increases.  Additionally, moving 
across spectral type may 
not only change the frequency of giant planets, but also their distributions 
of mass and semi-major axis.  The analysis of \citet{cumming} relied 
exclusively on solar-type (FGK) planet hosts, with the number of planets 
orbiting M stars being too small to draw any conclusions about a possible 
dependence of planet distributions on spectral type.  These issues cannot be 
well addressed with further null results; they require a large number of 
detected planets at intermediate ($\sim$5-20 AU) and large ($>$20 AU) 
separations to make statistically 
significant statements on long-period giant planet populations.

Since this paper was first submitted, the discovery of several planet 
candidates, via 
direct imaging, was announced; planets were detected around the three stars, 
all of A spectral type, HR 8799 \citep{hr8799b}, Fomalhaut \citep{fomalhautb}, 
and $\beta$ Pic \citep{betapicb}.  These exciting discoveries are consistent 
with the predictions of \citet{johnson}, as even though planets within 100 AU 
(like the three found around HR 8799) should be easier to detect around 
lower-mass stars (where the self-luminosity of the star is smaller), similar 
planets have not been found around stars of solar mass or smaller.  However, 
each one of these planetary 
systems is different from the other.  Moreover, all are around more massive 
stars than were analyzed here.  Since the full survey papers for each of these 
discoveries have not yet been published, it is difficult to incorporate these 
results into our analysis of Sun-like (and less massive) stars.



\section{Conclusions}

We have used Monte Carlo simulations to examine the null result from three 
direct imaging surveys \citep{elena,sdifinal,gdps} to set constraints on the 
population of extrasolar giant planets.  We use three commonly cited sets 
of planet models \citep{burrows,cond,fortney} in order to reach conclusions 
as broad as possible.  
Doubling the sample size, as expected, increased the strength of our null 
results. However, including better modeling for giant planets--using the 
stellar 
mass dependence of giant planet frequency of \citet{johnson}, and the 
core-accretion based luminosity models of \citet{fortney}--have actually 
loosened the constraints reported in \citet{sdifinalpab}.  
There is still some uncertainty, however, in which if any of these models of 
planet luminosity is correct; likely the truth may 
fall in between that of the optimistic ``hot start'' models and the somewhat 
pessimistic ``core accretion'' model.

With the COND models of \citet{cond}, a planet more massive than 4 M$_{Jup}$ 
is found around 20\% or less of FGKM stars in orbits between 8.1 and 911 AU, 
at 68\% confidence.  These limits become 7.4 to 863 AU for \citet{burrows}, 
and 25 to 557 AU for the models of 
\citet{fortney}.  At 95\% confidence, 4 M$_{Jup}$ (and larger) planets are 
found around fewer than 
20\% of stars between 22 and 507 AU, 21 and 479 AU, and 82 and 276 AU for 
the models of \citet{cond}, \citet{burrows}, and \citet{fortney}, 
respectively.

Using the power law distribution of \citet{cumming}, with 
index -0.61, the upper cut-off for the distribution of giant planets is found 
at 30 and 65 AU, at the 68\% and 95\% confidence levels, respectively, using 
the models of \citet{cond}.  With the models of \citet{burrows} these 
limits become 28 and 56 AU, and with the \citet{fortney} models they are 
83 and 182 AU.

When we apply the \citet{johnson} dependency of planet fraction on stellar 
mass, the M stars in our sample (where we achieve the greatest sensitivity 
to planets), are weighted down to account for their decreased likelihood of 
hosting a giant planet.  As a result, the improved null results cited above 
retreat to levels similar to those cited in \citet{sdifinalpab} and 
\citet{gdps}.  Given our 
results, fewer than 20\% of solar-type stars have a $>$4 M$_{Jup}$ planet 
between 13 and 849 AU at 68\% confidence with the \citet{cond} models (also 
13 and 805 AU for the models of \citet{burrows}, and 41 and 504 AU for the 
\citet{fortney} models).  At 95\% confidence, for the models of 
\citet{cond}, \citet{burrows}, and \citet{fortney}, fewer than 20\% of 
1 M$_{\sun}$ stars have 4 M$_{Jup}$ planets between 30 and 466 AU, 30 and 
440 AU, and 123 and 218 AU, respectively.

Applying the \citet{johnson} results to the \citet{cumming} model for 
semi-major axis distribution, giant planets cannot exist beyond 37 and 
82 AU for the \citet{cond} models at 68\% and 95\% confidence.  The 
68\% and 95\% confidence figures become 36 and 82 AU for the 
\citet{cond} models and 104 and 234 AU for the models of \citet{fortney}.  In 
general, the \citet{johnson} dependence of planet fraction on stellar mass 
makes direct imaging planet searches more difficult, as the stars most likely 
to harbor giant planets are also the most luminous, giving extreme contrast 
ratios between star and planet that impede planet detection.

We note that while the constraints on giant planet populations from 
this and other work have, for the first time, reached the equivalent of 
extrasolar ``Kuiper Belts,'' there is still a gap ($\sim$5 - $\sim$30 AU) 
between these results for FGKM stars and those of radial 
velocity surveys, which focus more on the inner solar system.  Delving into 
this unprobed region from the direct imaging side can be achieved two ways: 
increasing sensitivity to planets at small separations (achievable with 
dedicated planet finders using ``extreme'' adaptive optics, e.g. 
GPI \citep{gpi07} and VLT-SPHERE \citep{sphere08}), or 
with large-scale surveys to increase the sample size of target stars, such 
as the 500 hour Near-Infrared Coronagraphic Imager (NICI) survey at Gemini 
South (Chun et al. 2008).  Our technique can be applied to the results from 
any direct imaging survey for giant exoplanets, requiring only the target 
list and achieved contrast curves.  By building up the statistics of null 
results, it will be possible to more directly focus direct imaging efforts on 
where planets are most likely to exist, and create a fuller picture of the 
distribution of extrasolar giant planets.  Additionally, such an analysis 
helps to put the survey into context with respect to previous work.  There is 
also no limitation based on observation wavelength, even when target stars are 
observed by multiple surveys, since simulated planets are advanced in their 
orbits and compared to each contrast curve.  As such, it will be interesting 
in future work to consider the results from L and M band surveys currently 
being conducted (e.g. \citet{kasper07}).  \citet{cond} and \citet{burrows} 
predict planets with significantly lower contrasts to their parent stars at 
these longer wavelengths, and 
so the inclusion of results from such surveys could strengthen our null 
results, especially at large separations.

\acknowledgments

We are grateful to the anonymous referee for providing many helpful comments 
that improved the quality of this paper.  We thank Beth Biller for the 
publication and compilation of the SDI contrast curves, as 
well as a large amount of useful input in preparing these simulations.  We 
also thank Elena Masciadri for providing published contrast curves, and 
the authors of 
\citet{gdps} for the careful preparation and publication of their achieved 
contrasts and 
observational techniques.  We thank Eric Mamajek for a great deal of 
assistance in determining the ages of our target stars, as well as Michael 
Liu for providing the broadband-to-narrowband colors of low-mass objects. We 
thank 
Remi Soummer for the idea of presenting sensitivity to planets as a grid of 
mass and semi-major axis points, and we thank Daniel Apai for presenting the 
idea of constructing a grid of semi-major axis power law indices and 
cut-offs.  We also thank Rainer Lenzen, Thomas Henning, and Wolfgang Brandner 
for their important work in the original data gathering, and helpful comments 
over the course of the project.  
This work makes use of data from the European Southern Observatory, 
under Program 70.C - 0777D, 70.C - 0777E, 71.C-0029A, 74.C-0548, 74.C-0549, 
and 76.C-0094.  Observations reported here were obtained at the MMT 
Observatory, a joint facility of the University of Arizona and the 
Smithsonian Institution.  Based on observations obtained at the Gemini 
Observatory, which is operated by the Association of Universities for 
Research in Astronomy, Inc., under a cooperative agreement 
with the NSF on behalf of the Gemini partnership: the National Science 
Foundation (United States), the Science and Technology Facilities Council 
(United Kingdom), the National Research Council (Canada), CONICYT (Chile), 
the Australian Research Council (Australia), Minist\'erio da Ci\^encia e 
Tecnologia (Brazil) and SECYT (Argentina) This publication makes use of data 
products from the Two 
Micron All-Sky Survey, which is a joint project of the University of 
Massachusetts and the Infrared Processing and Analysis Center/California 
Institute of Technology, funded by the National Aeronautics and Space 
Administration and the National Science Foundation.  This research has made 
use of the SIMBAD database, operated at CDS, Strasburg, France.  
ELN is supported by a 
Michelson Fellowship, without which this work would not have been possible.  
LMC was supported by an NSF CAREER award and by NASA's 
Origin of Solar Systems program.  

\bibliographystyle{apj}
\bibliography{apj-jour,sdi_meta01}
\clearpage





\clearpage

%
%
%
%

\begin{deluxetable}{lccccccccc}
\tablecolumns{9}
\tablewidth{0pc}
\tabletypesize{\tiny}
\tablecaption{Target Stars}
\tablehead{
\colhead{Target} & \colhead{RA\tablenotemark{1}} & \colhead{Dec\tablenotemark{1}} & \colhead{Distance (pc)\tablenotemark{2}} & \colhead{Sp. Type} & \colhead{Age (Myr)} & \colhead{V\tablenotemark{1}} & \colhead{H\tablenotemark{3}} & \colhead{Ks\tablenotemark{3}} & \colhead{Obs. Mode\tablenotemark{4}}}
\startdata
\citet{sdifinal} \\
\hline
HIP 1481 & 00 18 26.1 & -63 28 39.0 & 40.95 & F8/G0V & 30 & 7.46 & 6.25 & 6.15 & VLT SDI \\
HD 8558 & 01 23 21.2 & -57 28 50.7 & 49.29 & G6V & 30 & 8.54 & 6.95 & 6.85 & VLT SDI \\
HD 9054 & 01 28 08.7 & -52 38 19.2 & 37.15 & K1V & 30 & 9.35 & 6.94 & 6.83 & VLT SDI \\
HIP 9141 & 01 57 48.9 & -21 54 05.0 & 42.35 & G3/G5V & 30 & 8.11 & 6.55 & 6.47 & VLT SDI \\
BD+05 378 & 02 41 25.9 & +05 59 18.4 & 40.54 & M0 & 12 & 10.20 & 7.23 & 7.07 & VLT SDI \\
HD 17925 & 02 52 32.1 & -12 46 11.0 & 10.38 & K1V & 200 & 6.05 & 4.23 & 4.17 & VLT SDI/GDPS \\
Eps Eri & 03 32 55.8 & -09 27 29.7 & 3.22 & K2V & 1100 & 3.73 & 1.88 & 1.78 & VLT SDI/GDPS \\
V577 Per A & 03 33 13.5 & +46 15 26.5 & 33.77 & G5IV/V & 70 & 8.35 & 6.46 & 6.37 & MMT SDI \\
GJ 174 & 04 41 18.9 & +20 54 05.4 & 13.49 & K3V & 280 & 7.98 & 5.31 & 5.15 & VLT SDI \\
GJ 182 & 04 59 34.8 & +01 47 00.7 & 26.67 & M1Ve & 12 & 10.10 & 6.45 & 6.26 & VLT SDI/Ks/GDPS \\
HIP 23309 & 05 00 47.1 & -57 15 25.5 & 26.26 & M0/1 & 12 & 10.09 & 6.43 & 6.24 & VLT SDI/Ks \\
AB Dor & 05 28 44.8 & -65 26 54.9 & 14.94 & K2Vk & 70 & 6.93 & 4.84 & 4.69 & VLT SDI \\
UY Pic & 05 36 56.8 & -47 57 52.9 & 23.87 & K0V & 70 & 7.95 & 5.93 & 5.81 & VLT SDI \\
AO Men & 06 18 28.2 & -72 02 41.4 & 38.48 & K6/7 & 12 & 10.99 & 6.98 & 6.81 & VLT SDI/Ks \\
HIP 30030 & 06 19 08.1 & -03 26 20.0 & 52.36 & G0V & 30 & 8.00 & 6.59 & 6.55 & MMT SDI \\
HIP 30034 & 06 19 12.9 & -58 03 16.0 & 45.52 & K2V & 30 & 9.10 & 7.09 & 6.98 & VLT SDI \\
HD 45270 & 06 22 30.9 & -60 13 07.1 & 23.50 & G1V & 70 & 6.50 & 5.16 & 5.05 & VLT SDI \\
HD 48189 A & 06 38 00.4 & -61 32 00.2 & 21.67 & G1/G2V & 70 & 6.15 & 4.75 & 4.54 & VLT SDI \\
pi01 UMa & 08 39 11.7 & +65 01 15.3 & 14.27 & G1.5V & 200 & 5.63 & 4.28 & 4.17 & MMT SDI/GDPS \\
HD 81040 & 09 23 47.1 & +20 21 52.0 & 32.56 & G0V & 2500 & 7.74 & 6.27 & 6.16 & MMT SDI \\
LQ Hya & 09 32 25.6 & -11 11 04.7 & 18.34 & K0V & 13 & 7.82 & 5.60 & 5.45 & MMT/VLT SDI/Ks/GDPS \\
DX Leo & 09 32 43.7 & +26 59 18.7 & 17.75 & K0V & 200 & 7.01 & 5.24 & 5.12 & MMT/VLT SDI/GDPS \\
HD 92945 & 10 43 28.3 & -29 03 51.4 & 21.57 & K1V & 70 & 7.76 & 5.77 & 5.66 & VLT SDI/GDPS \\
GJ 417 & 11 12 32.4 & +35 48 50.7 & 21.72 & G0V & 200 & 6.41 & 5.02 & 4.96 & MMT SDI/GDPS \\
TWA 14 & 11 13 26.5 & -45 23 43.0 & 46.00\tablenotemark{5} & M0 & 10 & 13.00 & 8.73 & 8.49 & VLT SDI \\
TWA 25 & 12 15 30.8 & -39 48 42.0 & 44.00\tablenotemark{5} & M0 & 10 & 11.40 & 7.50 & 7.31 & VLT SDI \\
RXJ1224.8-7503 & 12 24 47.3 & -75 03 09.4 & 24.17 & K2 & 16 & 10.51 & 7.84 & 7.71 & VLT SDI \\
HD 114613 & 13 12 03.2 & -37 48 10.9 & 20.48 & G3V & 8800 & 4.85 & 3.35 & 3.30 & VLT SDI \\
HD 128311 & 14 36 00.6 & +09 44 47.5 & 16.57 & K0 & 630 & 7.51 & 5.30 & 5.14 & MMT SDI \\
EK Dra & 14 39 00.2 & +64 17 30.0 & 33.94 & G0 & 70 & 7.60 & 6.01 & 5.91 & MMT SDI/GDPS \\
HD 135363 & 15 07 56.3 & +76 12 02.7 & 29.44 & G5V & 3 & 8.72 & 6.33 & 6.19 & MMT SDI/GDPS \\
KW Lup & 15 45 47.6 & -30 20 55.7 & 40.92 & K2V & 2 & 9.37 & 6.64 & 6.46 & VLT SDI \\
HD 155555 AB & 17 17 25.5 & -66 57 04.0 & 30.03 & G5IV & 12 & 7.20 & 4.91 & 4.70 & VLT SDI/Ks \\
HD 155555 C & 17 17 27.7 & -66 57 00.0 & 30.03 & M4.5 & 12 & 12.70 & 7.92 & 7.63 & VLT SDI/Ks \\
HD 166435 & 18 09 21.4 & +29 57 06.2 & 25.24 & G0 & 110 & 6.85 & 5.39 & 5.32 & MMT SDI \\
HD 172555 A\tablenotemark{6} & 18 45 26.9 & -64 52 16.5 & 29.23 & A5IV/V & 12 & 4.80 & 4.25 & 4.30 & VLT SDI \\
CD -64 1208 & 18 45 37.0 & -64 51 44.6 & 34.21 & K7 & 12 & 10.12 & 6.32 & 6.10 & VLT SDI/Ks \\
HD 181321 & 19 21 29.8 & -34 59 00.5 & 20.86 & G1/G2V & 160 & 6.48 & 5.05 & 4.93 & VLT SDI \\
HD 186704 & 19 45 57.3 & +04 14 54.6 & 30.26 & G0 & 210 & 7.03 & 5.62 & 5.52 & MMT SDI \\
GJ 799B & 20 41 51.1 & -32 26 09.0 & 10.22 & M4.5e & 12 & 11.00 & 5.20 & -99.00 & VLT SDI/Ks \\
GJ 799A & 20 41 51.2 & -32 26 06.6 & 10.22 & M4.5e & 12 & 10.25 & 5.20 & 4.94 & VLT SDI/Ks \\
GJ 803 & 20 45 09.5 & -31 20 27.1 & 9.94 & M0Ve & 12 & 8.81 & 4.83 & 4.53 & VLT SDI/Ks/GDPS \\
HD 201091 & 21 06 53.9 & +38 44 57.9 & 3.48 & K5Ve & 2000 & 5.21 & 2.54 & 2.25 & MMT SDI \\
Eps Indi A & 22 03 21.7 & -56 47 09.5 & 3.63 & K5Ve & 4000 & 4.69 & 2.35 & 2.24 & VLT SDI \\
GJ 862 & 22 29 15.2 & -30 01 06.4 & 15.45 & K5V & 6300 & 7.65 & 5.28 & 5.11 & VLT SDI \\
HIP 112312 A & 22 44 57.8 & -33 15 01.0 & 23.61 & M4e & 12 & 12.20 & 7.15 & 6.93 & VLT SDI \\
HD 224228 & 23 56 10.7 & -39 03 08.4 & 22.08 & K3V & 70 & 8.20 & 6.01 & 5.91 & VLT SDI \\
\hline
\citet{elena} \\
\hline
HIP 2729 & 00 34 51.2 & -61 54 58 & 45.91 & K5V & 30 & 9.56 & 6.72 & 6.53 & VLT Ks \\
BD +2 1729 & 07 39 23.0 & 02 11 01 & 14.87 & K7 & 200 & 9.82 & 6.09 & 5.87 & VLT H/GDPS \\
TWA 6 & 10 18 28.8 & -31 50 02 & 77.00\tablenotemark{5} & K7 & 10 & 11.62 & 8.18 & 8.04 & VLT Ks \\
BD +1 2447 & 10 28 55.5 & 00 50 28 & 7.23 & M2 & 70 & 9.63 & 5.61 & 5.31 & VLT H/GDPS \\
TWA 8A & 11 32 41.5 & -26 51 55 & 21.00\tablenotemark{5} & M2 & 10 & 12.10 & 7.66 & 7.43 & VLT Ks \\
TWA 8B & 11 32 41.5 & -26 51 55 & 21.00\tablenotemark{5} & M5 & 10 & 15.20 & 9.28 & 9.01 & VLT Ks \\
TWA 9A & 11 48 24.2 & -37 28 49 & 50.33 & K5 & 10 & 11.26 & 8.03 & 7.85 & VLT Ks \\
TWA 9B & 11 48 24.2 & -37 28 49 & 50.33 & M1 & 10 & 14.10 & 9.38 & 9.15 & VLT Ks \\
SAO 252852 & 14 42 28.1 & -64 58 43 & 16.40\tablenotemark{7} & K5V & 200 & 8.47 & 5.69 & 5.51 & VLT H \\
V343 Nor & 15 38 57.6 & -57 42 27 & 39.76 & K0V & 12 & 8.14 & 5.99 & 5.85 & VLT Ks \\
PZ Tel & 18 53 05.9 & -50 10 50 & 49.65 & K0Vp & 12 & 8.42 & 6.49 & 6.37 & VLT Ks \\
BD-17 6128 & 20 56 02.7 & -17 10 54 & 47.70 & K7 & 12 & 10.60 & 7.25 & 7.04 & VLT Ks \\
\hline
\citet{gdps} \\
\hline
HD 166 & 00 06 36.7839 & +29 01 17.406 & 13.70 & K0V & 200 & 6.13 & 4.63 & 4.31 & GDPS \\
HD 691 & 00 11 22.4380 & +30 26 58.470 & 34.10 & K0V & 260 & 7.96 & 6.26 & 6.18 & GDPS \\
HD 1405 & 00 18 20.890 & +30 57 22.23 & 30.60 & K2V & 70 & 8.60 & 6.51 & 6.39 & GDPS \\
HD 5996 & 01 02 57.2224 & +69 13 37.415 & 25.80 & G5V & 440 & 7.67 & 5.98 & 5.90 & GDPS \\
HD 9540 & 01 33 15.8087 & -24 10 40.662 & 19.50 & K0V & 2900 & 6.96 & 5.27 & 5.16 & GDPS \\
HD 10008 & 01 37 35.4661 & -06 45 37.525 & 23.60 & G5V & 200 & 7.66 & 5.90 & 5.75 & GDPS \\
HD 14802 & 02 22 32.5468 & -23 48 58.774 & 21.90 & G0V & 5200 & 5.19 & 3.71 & 3.74 & GDPS \\
HD 16765 & 02 41 13.9985 & -00 41 44.351 & 21.60 & F7IV & 290 & 5.71 & 4.64 & 4.51 & GDPS \\
HD 17190 & 02 46 15.2071 & +25 38 59.636 & 25.70 & K1IV & 4300 & 7.81 & 6.00 & 5.87 & GDPS \\
HD 17382 & 02 48 09.1429 & +27 04 07.075 & 22.40 & K1V & 430 & 7.62 & 5.69 & 5.61 & GDPS \\
HD 18803 & 03 02 26.0271 & +26 36 33.263 & 21.20 & G8V & 4400 & 6.72 & 5.02 & 4.95 & GDPS \\
HD 19994 & 03 12 46.4365 & -01 11 45.964 & 22.40 & F8V & 6200 & 5.06 & 3.77 & 3.75 & GDPS \\
HD 20367 & 03 17 40.0461 & +31 07 37.372 & 27.10 & G0V & 380 & 6.41 & 5.12 & 5.04 & GDPS \\
2E 759 & 03 20 49.50 & -19 16 10.0 & 27.00 & K7V & 200 & 10.26 & 7.66 & 7.53 & GDPS \\
HIP 17695 & 03 47 23.3451 & -01 58 19.927 & 16.30 & M3e & 70 & 11.59 & 7.17 & 6.93 & GDPS \\
HD 25457 & 04 02 36.7449 & -00 16 08.123 & 19.20 & F5V & 70 & 5.38 & 4.34 & 4.18 & GDPS \\
HD 283750 & 04 36 48.2425 & +27 07 55.897 & 17.90 & K2 & 300 & 8.42 & 5.40 & 5.24 & GDPS \\
HD 30652 & 04 49 50.4106 & +06 57 40.592 & 8.00 & F6V & 4500 & 3.19 & 1.76 & 1.60 & GDPS \\
HD 75332 & 08 50 32.2234 & +33 17 06.189 & 28.70 & F7V & 270 & 6.22 & 5.03 & 4.96 & GDPS \\
HD 77407 & 09 03 27.0820 & +37 50 27.520 & 30.10 & G0 & 120 & 7.10 & 5.53 & 5.44 & GDPS \\
HD 78141 & 09 07 18.0765 & +22 52 21.566 & 21.40 & K0 & 270 & 7.99 & 5.92 & 5.78 & GDPS \\
HD 90905 & 10 29 42.2296 & +01 29 28.025 & 31.60 & G0V & 230 & 6.90 & 5.60 & 5.52 & GDPS \\
HD 91901 & 10 36 30.7915 & -13 50 35.817 & 31.60 & K2V & 1000 & 8.75 & 6.64 & 6.57 & GDPS \\
HD 93528 & 10 47 31.1553 & -22 20 52.927 & 34.90 & K1V & 310 & 8.36 & 6.56 & 6.51 & GDPS \\
HIP 53020 & 10 50 52.0645 & +06 48 29.336 & 5.60 & M4 & 200 & 11.66 & 6.71 & 6.37 & GDPS \\
HD 96064 & 11 04 41.4733 & -04 13 15.924 & 24.60 & G8V & 250 & 8.41 & 5.90 & 5.80 & GDPS \\
HD 102392 & 11 47 03.8343 & -11 49 26.573 & 24.60 & K4.5V & 3400 & 9.05 & 6.36 & 6.19 & GDPS \\
HD 105631 & 12 09 37.2563 & +40 15 07.399 & 24.30 & K0V & 1500 & 8.26 & 5.70 & 5.60 & GDPS \\
HD 107146 & 12 19 06.5015 & +16 32 53.869 & 28.50 & G2V & 190 & 7.07 & 5.61 & 5.54 & GDPS \\
HD 108767 B & 12 29 50.908 & -16 31 14.99 & 26.90 & K2V & 140 & 8.51 & 6.37 & 6.24 & GDPS \\
HD 109085 & 12 32 04.2270 & -16 11 45.627 & 18.20 & F2V & 100 & 4.31 & 3.37 & 3.37 & GDPS \\
BD +60 1417 & 12 43 33.2724 & +60 00 52.656 & 17.70 & K0 & 270 & 9.40 & 7.36 & 7.29 & GDPS \\
HD 111395 & 12 48 47.0484 & +24 50 24.813 & 17.20 & G5V & 1000 & 6.31 & 4.70 & 4.64 & GDPS \\
HD 113449 & 13 03 49.6555 & -05 09 42.524 & 22.10 & K1V & 70 & 7.69 & 5.67 & 5.51 & GDPS \\
HD 116956 & 13 25 45.5321 & +56 58 13.776 & 21.90 & G9V & 710 & 7.29 & 5.48 & 5.41 & GDPS \\
HD 118100 & 13 34 43.2057 & -08 20 31.333 & 19.80 & K4.5V & 280 & 9.31 & 6.31 & 6.12 & GDPS \\
HD 124106 & 14 11 46.1709 & -12 36 42.358 & 23.10 & K1V & 1700 & 7.92 & 5.95 & 5.86 & GDPS \\
HD 130004 & 14 45 24.1821 & +13 50 46.734 & 19.50 & K2.5V & 5100 & 7.60 & 5.67 & 5.61 & GDPS \\
HD 130322 & 14 47 32.7269 & -00 16 53.314 & 29.80 & KOIII & 2900 & 8.05 & 6.32 & 6.23 & GDPS \\
HD 130948 & 14 50 15.8112 & +23 54 42.639 & 17.90 & G2V & 420 & 5.88 & 4.69 & 4.46 & GDPS \\
HD 139813 & 15 29 23.5924 & +80 27 00.961 & 21.70 & G5 & 270 & 7.31 & 5.56 & 5.45 & GDPS \\
HD 141272 & 15 48 09.4630 & +01 34 18.262 & 21.30 & G9V & 280 & 7.44 & 5.61 & 5.50 & GDPS \\
HIP 81084 & 16 33 41.6081 & -09 33 11.954 & 31.93 & K9Vkee & 70 & 11.29 & 7.78 & 7.55 & GDPS \\
HD 160934 & 17 38 39.6261 & +61 14 16.125 & 24.54 & K7 & 70 & 10.18 & 7.00 & 6.81 & GDPS \\
HD 166181 & 18 08 16.030 & +29 41 28.12 & 32.58 & G5V & 60 & 7.70 & 5.61 & 5.61 & GDPS \\
HD 167605 & 18 09 55.5001 & +69 40 49.788 & 30.96 & K2V & 500 & 8.60 & 6.45 & 6.33 & GDPS \\
HD 187748 & 19 48 15.4478 & +59 25 22.446 & 28.37 & G0 & 140 & 6.66 & 5.32 & 5.26 & GDPS \\
HD 201651 & 21 06 56.3893 & +69 40 28.548 & 32.84 & K0 & 6800 & 8.20 & 6.41 & 6.34 & GDPS \\
HD 202575 & 21 16 32.4674 & +09 23 37.772 & 16.17 & K3V & 700 & 7.91 & 5.53 & 5.39 & GDPS \\
HIP 106231 & 21 31 01.7137 & +23 20 07.374 & 25.06 & K3Vke & 70 & 9.24 & 6.52 & 6.38 & GDPS \\
HD 206860 & 21 44 31.3299 & +14 46 18.981 & 18.39 & G0VCH-0.5 & 200 & 6.00 & 4.60 & 4.56 & GDPS \\
HD 208313 & 21 54 45.0401 & +32 19 42.851 & 20.32 & K2V & 6400 & 7.78 & 5.68 & 5.59 & GDPS \\
V383 Lac & 22 20 07.0258 & +49 30 11.763 & 10.68 & K0 & 40 & 8.57 & 6.58 & 6.51 & GDPS \\
HD 213845 & 22 34 41.6369 & -20 42 29.577 & 22.74 & F5V & 200 & 5.20 & 4.27 & 4.33 & GDPS \\
HIP 114066 & 23 06 04.8428 & +63 55 34.359 & 24.94 & M0 & 70 & 10.87 & 7.17 & 6.98 & GDPS \\
HD 220140 & 23 19 26.6320 & +79 00 12.666 & 19.74 & K2Vk & 85 & 7.73 & 5.51 & 5.40 & GDPS \\
HD 221503 & 23 32 49.3999 & -16 50 44.307 & 13.95 & K6Vk & 550 & 8.60 & 5.61 & 5.47 & GDPS \\
HIP 117410 & 23 48 25.6931 & -12 59 14.849 & 27.06 & K5Vke & 55 & 9.57 & 6.49 & 6.29 & GDPS \\
\enddata
\tablenotetext{1}{from the CDS Simbad service}
\tablenotetext{2}{derived from the Hipparcos survey \citet{hip}}
\tablenotetext{3}{from the 2MASS Survey \citet{2mass}}
\tablenotetext{4}{In cases where target stars were observed by multiple surveys, the star is listed only in the first section of this table where it appears, either in the \citet{sdifinal} or \citet{elena} section, with Observing Mode given as ``VLT SDI/Ks'' or ``VLT H/GDPS,'' for example.}
\tablenotetext{5}{Distance from \citet{SZB03}}
\tablenotetext{6}{As this is the only star in our sample earlier than F2, we consider this work to be a survey of FGKM stars.}
\tablenotetext{7}{Distance from \citet{ZSBW01}}
\label{table1}
\end{deluxetable}

%
%
%
%

%
%
%
%
%

\begin{deluxetable}{lcccccccc}
\tablecolumns{8}
\tablewidth{0pc}
\tabletypesize{\tiny}
\tablecaption{Age Determination for Target Stars}
\tablehead{
\colhead{Target} & \colhead{Sp. Type\tablenotemark{*}} & \colhead{Li EW (mas)\tablenotemark{*}} & \colhead{Li Age (Myr)} & \colhead{R'$_{HK}$\tablenotemark{*}} & \colhead{R'$_{HK}$ Age\tablenotemark{++}} & \colhead{Group Membership\tablenotemark{1}} & \colhead{Group Age\tablenotemark{1}} & \colhead{Adopted Age}\tablenotemark{+++}}
\startdata
\citet{sdifinal} \\
\hline
HIP 1481 & F8/G0V\tablenotemark{2} & 129\tablenotemark{3} & 100 & -4.360\tablenotemark{4} & 221 & Tuc/Hor & 30 & 30 \\
HD 8558 & G6V\tablenotemark{2} & 205\tablenotemark{5} & 13 &   &   & Tuc/Hor & 30 & 30 \\
HD 9054 & K1V\tablenotemark{2} & 170\tablenotemark{5} & 160 & -4.236\tablenotemark{6} & 100 & Tuc/Hor & 30 & 30 \\
HIP 9141 & G3/G5V\tablenotemark{7} & 181\tablenotemark{8} & 13 &   &   & Tuc/Hor & 30 & 30 \\
BD+05 378 & M0\tablenotemark{9} &   &   &   &   & $\beta$ Pic & 12 & 12 \\
HD 17925 & K1V\tablenotemark{7} & 194\tablenotemark{8} & 50 & -4.357\tablenotemark{6} & 216 & Her/Lyr & 200 & 200 \\
Eps Eri & K2V\tablenotemark{10} &   &   & -4.598\tablenotemark{6} & 1129\tablenotemark{+} &   &   & 1100 \\
V577 Per A & G5IV/V\tablenotemark{11} & 219\tablenotemark{11} & 3 &   &   & AB Dor & 70 & 70 \\
GJ 174 & K3V\tablenotemark{12} & 45\tablenotemark{8} & 280 & -4.066\tablenotemark{13} &   &   &   & 280 \\
GJ 182 & M1Ve\tablenotemark{14} & 280\tablenotemark{15} & 12 &   &   &   &   & 12 \\
HIP 23309 & M0/1\tablenotemark{16} & 294\tablenotemark{16} & 12 & -3.893\tablenotemark{6} &   & $\beta$ Pic & 12 & 12 \\
AB Dor & K2Vk\tablenotemark{17} & 267\tablenotemark{8} & 10 & -3.880\tablenotemark{6} & $<$50 & AB Dor & 70 & 70 \\
UY Pic & K0V\tablenotemark{18} & 263\tablenotemark{8} & 10 & -4.234\tablenotemark{6} & 78 & AB Dor & 70 & 70 \\
AO Men & K6/7\tablenotemark{16} & 357\tablenotemark{16} & 6 & -3.755\tablenotemark{6} &   & $\beta$ Pic & 12 & 12 \\
HIP 30030 & G0V\tablenotemark{19} & 219\tablenotemark{8} & 2 &   &   & Tuc/Hor & 30 & 30 \\
HIP 30034 & K2V\tablenotemark{2} &   &   &   &   & Tuc/Hor & 30 & 30 \\
HD 45270 & G1V\tablenotemark{2} & 149\tablenotemark{5} & 90 & -4.378\tablenotemark{6} & 254 & AB Dor & 70 & 70 \\
HD 48189 A & G1/G2V\tablenotemark{2} & 145\tablenotemark{8} & 25 & -4.268\tablenotemark{6} & 105 & AB Dor & 70 & 70 \\
pi01 UMa & G1.5V\tablenotemark{20} & 135\tablenotemark{8} & 100 & -4.400\tablenotemark{21} & 300 &   &   & 200 \\
HD 81040 & G0V\tablenotemark{20} & 24\tablenotemark{22} & 2500 &   &   &   &   & 2500 \\
LQ Hya & K0V\tablenotemark{20} & 247\tablenotemark{8} & 13 &   &   &   &   & 13 \\
DX Leo & K0V\tablenotemark{20} & 180\tablenotemark{8} & 100 & -4.234\tablenotemark{6} & 78 & Her/Lyr & 200 & 200 \\
HD 92945 & K1V\tablenotemark{20} & 138\tablenotemark{8} & 160 & -4.393\tablenotemark{6} & 285 & AB Dor & 70 & 70 \\
GJ 417 & G0V\tablenotemark{23} & 76\tablenotemark{24} & 250 & -4.368\tablenotemark{13} & 235 & Her/Lyr & 200 & 200 \\
TWA 14 & M0\tablenotemark{25} & 600\tablenotemark{25} & 8 &   &   & TW Hya & 10 & 10 \\
TWA 25 & M0\tablenotemark{9} & 494\tablenotemark{26} & 10 &   &   & TW Hya & 10 & 10 \\
RXJ1224.8-7503 & K2\tablenotemark{27} & 250\tablenotemark{27} & 16 &   &   &   &   & 16 \\
HD 114613 & G3V\tablenotemark{28} & 100\tablenotemark{29} & 400 & -5.118\tablenotemark{6} & 7900 &   &   & 8800 \\
HD 128311 & K0\tablenotemark{20} &   &   & -4.489\tablenotemark{13} & 565 &   &   & 630 \\
EK Dra & G0\tablenotemark{30} & 212\tablenotemark{8} & 2 & -4.106\tablenotemark{13} & $<$50 & AB Dor & 70 & 70 \\
HD 135363 & G5V\tablenotemark{20} & 220\tablenotemark{8} & 3 &   &   &   &   & 3 \\
KW Lup & K2V\tablenotemark{28} & 430\tablenotemark{31} & 2 &   &   &   &   & 2 \\
HD 155555 AB & G5IV\tablenotemark{16} & 205\tablenotemark{8} & 6 & -3.965\tablenotemark{6} & $<$50 & $\beta$ Pic & 12 & 12 \\
HD 155555 C & M4.5\tablenotemark{16} &   &   &   &   & $\beta$ Pic & 12 & 12 \\
HD 166435 & G0\tablenotemark{32} &   &   & -4.270\tablenotemark{21} & 107 &   &   & 110 \\
HD 172555 A & A5IV/V\tablenotemark{2} &   &   &   &   & $\beta$ Pic & 12 & 12 \\
CD -64 1208 & K7\tablenotemark{16} & 580\tablenotemark{16} & 5 &   &   & $\beta$ Pic & 12 & 12 \\
HD 181321 & G1/G2V\tablenotemark{28} & 131\tablenotemark{8} & 79 & -4.372\tablenotemark{6} & 243 &   &   & 160 \\
HD 186704 & G0\tablenotemark{33} &   &   & -4.350\tablenotemark{21} & 205 &   &   & 210 \\
GJ 799B & M4.5e\tablenotemark{34} &   &   &   &   & $\beta$ Pic & 12 & 12 \\
GJ 799A & M4.5e\tablenotemark{34} &   &   &   &   & $\beta$ Pic & 12 & 12 \\
GJ 803 & M0Ve\tablenotemark{34} & 51\tablenotemark{8} & 30 &   &   & $\beta$ Pic & 12 & 12 \\
HD 201091 & K5Ve\tablenotemark{34} &   &   & -4.704\tablenotemark{13} & 2029\tablenotemark{+} &   &   & 2000 \\
Eps Indi A & K5Ve\tablenotemark{34} &   &   & -4.851\tablenotemark{6} & 3964\tablenotemark{+} &   &   & 4000 \\
GJ 862 & K5V\tablenotemark{34} &   &   & -4.983\tablenotemark{6} & 6280\tablenotemark{+} &   &   & 6300 \\
HIP 112312 A & M4e\tablenotemark{9} &   &   &   &   & $\beta$ Pic & 12 & 12 \\
HD 224228 & K3V\tablenotemark{28} & 53\tablenotemark{8} & 630 & -4.468\tablenotemark{6} &   & AB Dor & 70 & 70 \\
\hline
\citet{elena} \\
\hline
HIP 2729 & K5V\tablenotemark{2} &   &   &   &   & Tuc/Hor & 30 & 30 \\
BD +2 1729 & K7\tablenotemark{20} &   &   &   &   & Her/Lyr & 200 & 200 \\
TWA 6 & K7\tablenotemark{35} & 560\tablenotemark{35} & 3 &   &   & TW Hya & 10 & 10 \\
BD +1 2447 & M2\tablenotemark{36} &   &   &   &   & AB Dor & 70 & 70 \\
TWA 8A & M2\tablenotemark{35} & 530\tablenotemark{35} & 3 &   &   & TW Hya & 10 & 10 \\
TWA 8B & M5\tablenotemark{35} & 560\tablenotemark{35} & 3 &   &   & TW Hya & 10 & 10 \\
TWA 9A & K5\tablenotemark{35} & 460\tablenotemark{35} & 3 &   &   & TW Hya & 10 & 10 \\
TWA 9B & M1\tablenotemark{35} & 480\tablenotemark{35} & 3 &   &   & TW Hya & 10 & 10 \\
SAO 252852 & K5V\tablenotemark{37} &   &   &   &   & Her/Lyr & 200 & 200 \\
V343 Nor & K0V\tablenotemark{2} & 300\tablenotemark{29} & 5 & -4.159\tablenotemark{6} & 40 & $\beta$ Pic & 12 & 12 \\
PZ Tel & K0Vp\tablenotemark{18} & 267\tablenotemark{38} & 20 & -3.780\tablenotemark{4} & $<$50 & $\beta$ Pic & 12 & 12 \\
BD-17 6128 & K7\tablenotemark{39} & 400\tablenotemark{40} & 3 &   &   & $\beta$ Pic & 12 & 12 \\
\hline
\citet{gdps} \\
\hline
HD 166 & K0V\tablenotemark{41} & 74\tablenotemark{42} & 290 & -4.458\tablenotemark{13} & 460 & Her/Lyr & 200 & 200 \\
HD 691 & K0V\tablenotemark{43} & 110\tablenotemark{8} & 260 & -4.380\tablenotemark{21} & 260 &   &   & 260 \\
HD 1405 & K2V\tablenotemark{44} & 271\tablenotemark{45} &   &   &   & AB Dor & 70 & 70 \\
HD 5996 & G5V\tablenotemark{46} &   &   & -4.454\tablenotemark{13} & 440 &   &   & 440 \\
HD 9540 & K0V\tablenotemark{7} &   &   & -4.774\tablenotemark{6} & 2900 &   &   & 2900 \\
HD 10008 & G5V\tablenotemark{47} & 103\tablenotemark{48} & 280 & -4.530\tablenotemark{13} & 740 & Her/Lyr & 200 & 200 \\
HD 14802 & G0V\tablenotemark{17} & 51\tablenotemark{49} & 4000 & -4.985\tablenotemark{6} & 6300 &   &   & 5200 \\
HD 16765 & F7IV\tablenotemark{50} & 73\tablenotemark{14} & 270 & -4.400\tablenotemark{13} & 300 &   &   & 290 \\
HD 17190 & K1IV\tablenotemark{51} &   &   & -4.870\tablenotemark{21} & 4300 &   &   & 4300 \\
HD 17382 & K1V\tablenotemark{51} &   &   & -4.450\tablenotemark{21} & 430 &   &   & 430 \\
HD 18803 & G8V\tablenotemark{52} &   &   & -4.880\tablenotemark{21} & 4400 &   &   & 4400 \\
HD 19994 & F8V\tablenotemark{53} & 12\tablenotemark{54} & 8000 & -4.880\tablenotemark{21} & 4400 &   &   & 6200 \\
HD 20367 & G0V\tablenotemark{55} & 113\tablenotemark{8} & 150 & -4.500\tablenotemark{21} & 610 &   &   & 380 \\
2E 759 & K7V\tablenotemark{56} & 63\tablenotemark{57} & 260 &   &   & Her/Lyr & 200 & 200 \\
HIP 17695 & M3e\tablenotemark{58} &   &   &   &   & AB Dor & 70 & 70 \\
HD 25457 & F5V\tablenotemark{59} & 91\tablenotemark{60} & 80 & -4.390\tablenotemark{21} & 280 & AB Dor & 70 & 70 \\
HD 283750 & K2\tablenotemark{61} & 33\tablenotemark{8} & 300 & -4.057\tablenotemark{13} &   &   &   & 300 \\
HD 30652 & F6V\tablenotemark{62} & 15\tablenotemark{14} & 7500 & -4.650\tablenotemark{21} & 1500 &   &   & 4500 \\
HD 75332 & F7V\tablenotemark{50} & 125\tablenotemark{8} & 50 & -4.470\tablenotemark{21} & 500 &   &   & 270 \\
HD 77407 & G0\tablenotemark{63} & 162\tablenotemark{45} & 50 & -4.340\tablenotemark{21} & 190 &   &   & 120 \\
HD 78141 & K0\tablenotemark{64} & 107\tablenotemark{8} & 270 &   &   &   &   & 270 \\
HD 90905 & G0V\tablenotemark{65} & 136\tablenotemark{8} & 80 & -4.430\tablenotemark{21} & 370 &   &   & 230 \\
HD 91901 & K2V\tablenotemark{7} & 7\tablenotemark{66} & 1000 &   &   &   &   & 1000 \\
HD 93528 & K1V\tablenotemark{17} & 100\tablenotemark{8} & 260 & -4.424\tablenotemark{6} & 360 &   &   & 310 \\
HIP 53020 & M4\tablenotemark{67} &   &   &   &   & Her/Lyr & 200 & 200 \\
HD 96064 & G8V\tablenotemark{68} & 114\tablenotemark{8} & 250 & -4.373\tablenotemark{13} & 250 &   &   & 250 \\
HD 102392 & K4.5V\tablenotemark{17} &   &   & -4.811\tablenotemark{6} & 3400\tablenotemark{+} &   &   & 3400 \\
HD 105631 & K0V\tablenotemark{69} &   &   & -4.650\tablenotemark{21} & 1500 &   &   & 1500 \\
HD 107146 & G2V\tablenotemark{70} & 125\tablenotemark{8} & 180 & -4.340\tablenotemark{21} & 190 &   &   & 190 \\
HD 108767 B & K2V\tablenotemark{71} & 175\tablenotemark{72} & 140 &   &   &   &   & 140 \\
HD 109085 & F2V\tablenotemark{17} & 37\tablenotemark{73} & 100 &   &   &   &   & 100 \\
BD +60 1417 & K0\tablenotemark{74} & 96\tablenotemark{8} & 270 &   &   &   &   & 270 \\
HD 111395 & G5V\tablenotemark{52} &   &   & -4.580\tablenotemark{21} & 1000 &   &   & 1000 \\
HD 113449 & K1V\tablenotemark{17} & 142\tablenotemark{8} & 200 & -4.340\tablenotemark{13} & 190 & AB Dor & 70 & 70 \\
HD 116956 & G9V\tablenotemark{68} & 31\tablenotemark{24} & 1000 & -4.447\tablenotemark{13} & 420 &   &   & 710 \\
HD 118100 & K4.5V\tablenotemark{68} & 25\tablenotemark{15} & 280 & -4.090\tablenotemark{13} &   &   &   & 280 \\
HD 124106 & K1V\tablenotemark{17} &   &   & -4.675\tablenotemark{6} & 1700 &   &   & 1700 \\
HD 130004 & K2.5V\tablenotemark{68} &   &   & -4.919\tablenotemark{13} & 5100\tablenotemark{+} &   &   & 5100 \\
HD 130322 & KOIII\tablenotemark{75} &   &   & -4.780\tablenotemark{21} & 2900 &   &   & 2900 \\
HD 130948 & G2V\tablenotemark{52} & 116\tablenotemark{8} & 230 & -4.500\tablenotemark{21} & 610 &   &   & 420 \\
HD 139813 & G5\tablenotemark{76} & 119\tablenotemark{8} & 230 & -4.400\tablenotemark{21} & 300 &   &   & 270 \\
HD 141272 & G9V\tablenotemark{68} &   &   & -4.390\tablenotemark{21} & 280 &   &   & 280 \\
HIP 81084 & K9Vkee\tablenotemark{77} &   &   & -4.210\tablenotemark{6} &   & AB Dor & 70 & 70 \\
HD 160934 & K7\tablenotemark{78} & 40\tablenotemark{79} & 280 &   &   & AB Dor & 70 & 70 \\
HD 166181 & G5V\tablenotemark{80} & 186\tablenotemark{8} & 60 &   &   &   &   & 60 \\
HD 167605 & K2V\tablenotemark{81} & 14\tablenotemark{57} & 500 &   &   &   &   & 500 \\
HD 187748 & G0\tablenotemark{63} & 114\tablenotemark{8} & 140 &   &   &   &   & 140 \\
HD 201651 & K0\tablenotemark{47} &   &   & -5.010\tablenotemark{21} & 6800 &   &   & 6800 \\
HD 202575 & K3V\tablenotemark{68} &   &   & -4.522\tablenotemark{13} & 700\tablenotemark{+} &   &   & 700 \\
HIP 106231 & K3Vke\tablenotemark{68} & 140\tablenotemark{8} & 180 & -3.906\tablenotemark{13} &   & AB Dor & 70 & 70 \\
HD 206860 & G0VCH-0.5\tablenotemark{17} & 110\tablenotemark{82} & 190 & -4.400\tablenotemark{6} & 300 & Her/Lyr & 200 & 200 \\
HD 208313 & K2V\tablenotemark{68} &   &   & -4.987\tablenotemark{13} & 6400\tablenotemark{+} &   &   & 6400 \\
V383 Lac & K0\tablenotemark{83} & 259\tablenotemark{8} & 40 &   &   &   &   & 40 \\
HD 213845 & F5V\tablenotemark{17} &   &   & -4.547\tablenotemark{6} & 830 & Her/Lyr & 200 & 200 \\
HIP 114066 & M0\tablenotemark{78} &   &   &   &   & AB Dor & 70 & 70 \\
HD 220140 & K2Vk\tablenotemark{68} & 218\tablenotemark{8} & 85 & -4.074\tablenotemark{13} &   &   &   & 85 \\
HD 221503 & K6Vk\tablenotemark{17} &   &   & -4.486\tablenotemark{6} & 550\tablenotemark{+} &   &   & 550 \\
HIP 117410 & K5Vke\tablenotemark{17} &   &   & -4.194\tablenotemark{6} & 55\tablenotemark{+} &   &   & 55 \\
\enddata
\tablenotetext{1}{Group Membership for TWA, $\beta$ Pic, Tuc/Hor, and AB Dor from \citet{ZS04}, Her/Lyr from \citet{LMCF06}.  Group Ages from \citet{ZS04} (TWA, $\beta$ Pic, and Tuc/Hor), \citet{abdorme} (AB Dor), and \citet{LMCF06} (Her/Lyr)}
\tablenotetext{*}{Measurement References: 2: \citet{HC75},  3: \citet{WCMM05},  4: \citet{Henry96},  5: \citet{TDQDJ00},  6: \citet{Gray06},  7: \citet{HS88},  8: \citet{WSH03},  9: \citet{ZS04},  10: \citet{CHW67},  11: \citet{CM02},  12: \citet{LP60},  13: \citet{Gray03},  14: \citet{FBMS95},  15: \citet{FMS97},  16: \citet{ZSBW01},  17: \citet{GCGMB06},  18: \citet{H78},  19: \citet{CPKR95},  20: \citet{MLGFDC01},  21: \citet{Wright04},  22: \citet{SUZT06},  23: \citet{B51},  24: \citet{GHH00},  25: \citet{ZWSB01},  26: \citet{SZB03},  27: \citet{AKSCWM95},  28: \citet{H82},  29: \citet{RGP93},  30: \citet{GJ79},  31: \citet{NB98},  32: \citet{HDC},  33: \citet{A85},  34: \citet{GJ91},  35: \citet{WZPPWSM99},  36: \citet{VJMW46},  37: \citet{E61},  38: \citet{SKH98},  39: \citet{NKAVB95},  40: \citet{MDCKDS95},  41: \citet{RB82},  42: \citet{ZBR97},  43: \citet{E62},  44: \citet{ABF98},  45: \citet{MLFG01},  46: \citet{HFET83},  47: \citet{hip},  48: \citet{LMCF06},  49: \citet{PLP94},  50: \citet{C76},  51: \citet{H56},  52: \citet{HT70},  53: \citet{HS55},  54: \citet{ISMR04},  55: \citet{SK90},  56: \citet{FGM89},  57: \citet{FBMS93},  58: \citet{ATZKV98},  59: \citet{M75},  60: \citet{LR04},  61: \citet{OHL88},  62: \citet{MK73},  63: \citet{P69},  64: \citet{SHLSBN00},  65: \citet{H74},  66: \citet{SWGSW00},  67: \citet{B85},  68: \citet{GCGMR03},  69: \citet{S73},  70: \citet{HT70e},  71: \citet{MMSM01},  72: \citet{PPR92},  73: \citet{MPP03},  74: \citet{RB88},  75: \citet{US70},  76: \citet{PMABB95},  77: \citet{FSBWG06},  78: \citet{RHG95},  79: \citet{ZSB04},  80: \citet{E64},  81: \citet{SMGMS91},  82: \citet{CNBZ01},  83: \citet{BLLWJM96}}
\tablenotetext{+}{In general, we have only determined Ca R'$_{HK}$ ages for stars with spectral types K1 or earlier, but in the case of these K2-K6 stars, we have only the R'$_{HK}$ measurement on which to rely for age determination.  The calibration of Mt. Wilson  S-index to R'$_{HK}$ for K5 stars (B-V $\sim$ 1.1 mag) has not been well-defined (\citet{noyes}; specifically the photospheric subtraction), and hence applying a R'$_{HK}$ vs. age relation for K5 stars is unlikely to yield accurate ages.}
\tablenotetext{++}{Using Eq. 3 of \citet{Mamajek08} to convert R'$_{HK}$ into age}
\tablenotetext{+++}{In general, ages derived from lithium and/or calcium alone are likely accurate to within a factor of $\sim$2}
\label{table2}
\end{deluxetable}

%
%
%
%

\begin{deluxetable}{lccccc}
\tablecolumns{4}
\tablewidth{0pc}
\tabletypesize{\tiny}
\tablecaption{Summary of Results.\label{table3a}}
\tablehead{
\colhead{Target Stars} & \colhead{Mass Correction\tablenotemark{*}} & \colhead{Confidence Level} & \colhead{\citet{cond}} & \colhead{\citet{burrows}} & \colhead{\citet{fortney}}}

\startdata
\multicolumn{6}{c}{Completeness plots: semi-major axis range with f${_p}<$ 20\% for M $>$ 4 M$_{Jup}$}\\
\hline
All & None & 68\% & 8.1 - 911 AU & 7.4 - 863 AU & 25 - 557 AU \\
All & None & 95\% & 22 - 507 AU & 21 - 479 AU & 82 - 276 AU \\
M stars & None & 68\% & 9.0 - 207 AU & 8.3 - 213 AU & 43 - 88 AU \\
M stars & None & 95\% & -- & -- & -- \\
FGK stars & None & 68\% & 25 - 856 AU & 25 - 807 AU & 59 - 497 AU \\
FGK stars & None & 95\% & 38 - 469 AU & 40 - 440 AU & -- \\
All & 1 M$_{\sun}$ & 68\% & 13 - 849 AU & 13 - 805 AU & 41 - 504 AU \\
All & 1 M$_{\sun}$ & 95\% & 30 - 466 AU & 30 - 440 AU & 123 - 218 AU \\
All & 0.5 M$_{\sun}$ & 68\% & 9.0 - 1070 AU & 8.3 - 1016 AU & 26 - 656 AU \\
All & 0.5 M$_{\sun}$ & 95\% & 23 - 605 AU & 22 - 573 AU & 71 - 341 AU \\
\hline
\multicolumn{6}{c}{Upper cut-off on power law distribution for semi-major axis with index -0.61}\\
\hline
All & None & 68\% & 30 AU & 28 AU & 83 AU \\
All & None & 95\% & 65 AU & 56 AU & 182 AU \\
All & Yes & 68\% & 37 AU & 36 AU & 104 AU \\
All & Yes & 95\% & 82 AU & 82 AU & 234 AU \\

\enddata

\tablenotetext{*}{The ``Mass Correction'' column refers to 
whether or not the \citet{johnson} result, that more massive stars are 
more likely to harbor giant planets, is used to weight the target stars by 
stellar mass.  For the completeness plots, this correction is either 
not applied (None) or set to a specific stellar mass, to determine the upper 
limit on the frequency of giant planets around stars of that mass.  For 
the limits on the upper cut-off on power law distributions, the correction 
is either applied (Yes) or not (None).}
\end{deluxetable}

\begin{deluxetable}{lcccc}
\tablecolumns{4}
\tablewidth{0pc}
\tabletypesize{\tiny}
\tablecaption{Binaries}
\tablehead{
\colhead{Target} & \colhead{Sep (``)\tablenotemark{1}} & \colhead{Sep. (AU)\tablenotemark{1}} & \colhead{Reference} & \colhead{Companion Type}}
\startdata
\citet{sdifinal} \\
\hline
HIP 9141 & 0.15 & 6.38 & \citet{sdifinal} & mid-G \\
V577 Per A & 7 & 230 & \citet{PABBB93} & M0 \\
AB Dor & 9 (Ba/Bb) & 134 (Ba/Bb) & \citet{abdor} & Binary M stars\\
AB Dor & 0.15 (C) & 2.24 (C) & \citet{abdor} & Very low-mass M Star\\
HIP 30034 & 5.5 & 250 & \citet{abpic} & Planet/Brown Dwarf \\
HD 48189 A & 0.76 (B) & 16.5 & \citet{FM00} & K star \\
HD 48189 A & 0.14 & 3.03 & \citet{sdifinal} & K star \\
DX Leo & 65 & 1200 & \citet{LBSKW05} & M5.5 \\
EK Dra & SB & SB & \citet{MH06} & M2\\
HD 135363 & 0.26 & 7.65 & \citet{sdifinal} & late K/early M\\
HD 155555 AB & SB (AB) & SB (AB) & \citet{BEL67} & G5 and K0 SB \\
HD 155555 AB & 18 (C) & 1060 (C) & \citet{ZSBW01} & Target Star 155555 C, M4.5\\
HD 172555 A & 71 & 2100 & \citet{SD93} & Target Star CD -64 1208, K7 \\
HD 186704 & 13 & 380 & \citet{A32} & early M\\
GJ 799A & 3.6 & 36 & \citet{W52} & Target Star GJ 799B, M4.5 \\
HD 201091 & 16 & 55 & \citet{B50} & K5 \\
Eps Indi A & 400 & 1500 & \citet{epsindi} & Binary Brown Dwarf \\
HIP 112312 & 100 & 2400 & \citet{SBZ02} & M4.5 \\

\hline
\citet{elena} \\
\hline

TWA 8A & 13 & 270 & \citet{JHFF99} & Target Star TWA 8B, M5 \\
TWA 9A & 9 & 576 & \citet{JHFF99} & Target Star TWA 9B, M1 \\
SAO 252852 & 15.7 & 260 & \citet{poveda} & HD 128898, Ap \\
V343 Nor & 10 & 432 & \citet{SZB03} & M4.5 \\
BD-17 6128 & 2 & 100 & \citet{NGMGE02} & M2\\

\hline
\citet{gdps} \\
\hline

HD 14802 & 0.47 & 10 & \citet{gdps} & ~K6 \\
HD 16765   & 4.14  & 89    & \citet{hol77}  & $\sim$K   \\
HD 17382   & 20.3    & 456   & \citet{lepine05}  & M4.5   \\
HD 19994   & $\sim$5    & $\sim$100   & \citet{hale94}  & M3V   \\
HD283750   & 124    & 2220   & \citet{holberg02}  & White Dwarf   \\
HD 77407   & 1.7    & 50   & \citet{mug04}  & $\sim$M   \\
HD 93528   & 234    & 8200   & \citet{hip}  & HIP 52776, K4.5\tablenotemark{2}   \\
HD 96064   & 11.47    & 283 AU   & \citet{lippi79}  & NLTT 26194, M3   \\
HD 102392   & 1.13    & 28   & \citet{gdps}  & $\sim$M   \\
HD 108767 B   & 23.7  & 639   & \citet{gould04}  & A0IV   \\
HD 130948   & 2.64    & 47   & \citet{potter02}  & Binary brown dwarfs   \\
HD 139813   & 31.5    & 683   & \citet{steph60}  & G0   \\
HD 141272   & 17.8    & 350   & \citet{eisen07}  & M3   \\
HD 160934   & SB    & SB   & \citet{hormuth07}  & early M, $a=$ 4.5 AU   \\
HD 160934   & 8.69    & 213   & \citet{LBSKW05}  & $\sim$M   \\
HD 166181   & SB    & SB   & \citet{nadal74}  & 1.8 day orbit   \\
HD 166181   & 0.102    & 3   & \citet{gdps}  &  $\sim$K  \\
HD 167605   & 1.2    & 37   & \citet{arribas98}  & M4V   \\
HD 206860   & 43.2    & 795   & \citet{luhman07}  & T dwarf   \\
HD 213845   & 6.09    & 139   & \citet{gdps}  & late M   \\
HD 220140   & 10.9    & 214   & \citet{LBSKW05}  & mid M   \\
HD 220140   & 963    &  19000  & \citet{makarov07}  & $\sim$M   \\
HD 221503   & 339   & 4700   & \citet{gould04}  & binary M stars   \\
HIP 117410   & 1.84    & 50   & \citet{ross55}  &  early M  \\

\enddata
\tablenotetext{1}{SB indicates a spectroscopic binary}
\tablenotetext{2}{These stars have \textit{Hipparcos} proper motion and parallax within errors, and similar values of calcium R'$_{HK}$ (-4.424 and -4.451 for HD 93528 and HIP 52776, respectively).}
\label{table3}
\end{deluxetable}

\end{document}